\pdfoutput=1
\documentclass[%
 reprint,
 amsmath,amssymb,longbibliography,
 aps,
 prb,
]{revtex4-2}%
\usepackage{hyperref}
\usepackage{multirow}
\usepackage{booktabs}
\usepackage[dvipsnames]{xcolor}
\usepackage{dsfont}
\usepackage{nicefrac}
\usepackage{bm}
\usepackage{graphicx}
\usepackage{amsmath}
\usepackage{color}

\usepackage{tikz}

\newcommand{\sruo}{Sr$_2$RuO$_4$}

\newcommand{\cd}{c^{\dagger}}
\usetikzlibrary{arrows,positioning} 
\tikzset{
    photon/.style={decorate, decoration={snake}, draw=black},
    electron/.style={draw=black, postaction={decorate},
        decoration={markings,mark=at position .55 with {\arrow[draw=black,thick]{>}}}},
    gluon/.style={decorate, decoration={snake},draw=black}, 
    >=stealth',
    punkt/.style={
           rectangle,
           rounded corners,
           draw=black, very thick,
           text width=6.5em,
           minimum height=2em,
           text centered},
    pil/.style={
           ->,
           thick,
           shorten <=2pt,
           shorten >=2pt,}
}

\begin{document}

\title{Supercurrents and spontaneous time-reversal symmetry breaking by nonmagnetic disorder in unconventional superconductors}

\author{Clara N. Brei\o$^1$, P. J. Hirschfeld$^2$, and Brian M. Andersen$^1$}
\affiliation{%
$^1$Niels Bohr Institute, University of Copenhagen, R\aa dmandsgade 62, DK-2200 Copenhagen, Denmark\\
$^2$Department of Physics, University of Florida, Gainesville, Florida 32611, USA
}

\date{\today}

\begin{abstract}
  Recently, a theoretical study [Li {\it et al.},~npj Quantum Materials 6, 36  (2021)] investigated a model of a disordered $d$-wave superconductor, and reported local time-reversal symmetry breaking current loops for sufficiently high disorder levels. Since the pure $d$-wave superconducting state does not break time-reversal symmetry, it is surprising that such persistent currents arise purely from nonmagnetic disorder. Here we perform a detailed theoretical investigation of such disorder-induced orbital currents, and show that the occurrence of the currents can be traced to local extended $s$-wave pairing. We discuss the energetics leading to regions of $s\pm id$ order, which support spontaneous  local currents in the presence of inhomogeneous density modulations.
\end{abstract}

\maketitle

\section{Introduction}

One of the fascinating possibilities currently under debate in the field of unconventional superconductivity is how to realize a superconducting state where time-reversal is spontaneously broken.  The possibility of doing so intrinsically has been known for some time in the form of the superfluid $^3$He-A phase \cite{VollhardtWoelfle}, where the orbital part of the pair wave function is chiral and is sometimes referred to as $p_x+ip_y$, or $p+ip$ for short.  Solid state $p+ip$ superconducting analogs have been proposed frequently, including for Sr$_2$RuO$_4$ \cite{Mackenzie2017}, UPt$_3$ \cite{Sauls2018}, and the 5/2 fractional quantum Hall state\cite{MooreRead1991}.

In the Sr$_2$RuO$_4$ case, recent NMR evidence argues strongly against spin-triplet $p+ip$ pairing \cite{Pustogow19,Chronister}, but $\mu$SR\cite{Luke1998} and Kerr\cite{Kapitulnik09} effect measurements support time-reversal symmetry breaking (TRSB), leading to suggestions of linear combinations of dominant spin-singlet complex combinations of two-dimensional (2D) irreducible representations  $d_{xz}+id_{yz}$ \cite{Pustogow19}, or combinations of various one-dimensional (1D) irreducible representations like $s+id_{x^2-y^2}$ \cite{RomerPRL,Romer_MPLB}, $s+id_{xy}$ \cite{Romer2021_Longerrange,clepkens2021}, or $d_{x^2-y^2}+ig$ \cite{kivelson2020proposal}. The $p+ip$ and $d+id$ states are chiral, and should support spontaneous edge currents~\cite{Karmakar2021}, while the remaining states break time-reversal symmetry (TRS), but are not chiral and so should not  support such  states.  This distinction is reflected in the behavior of the different TRSB states in their local response to nonmagnetic disorder.  A single impurity in a chiral state supports a net circulating spontaneous current that decays over long distances, whereas the same impurity in one of the non-chiral states creates a pattern of local currents that averages to zero over the sample \cite{Lee2009,Maiti2015,Lin2016}.   

With this intuition, it is almost trivial to deduce that a nonmagnetic impurity
in a superconducting state that does {\it not} break TRS will not create any spontaneous currents at all, and this is indeed found to be the case in simple models of, e.g. $d_{x^2-y^2}$ superconductors.  Recently, however, Li {\it et al.}\cite{Li2021} showed within Bogoliubov-de Gennes mean field simulations of high disorder levels in a $d$-wave superconductor, that local currents were generated.
Left open were questions of the origin of this local TRSB effect, in particular how multiple impurities can create TRSB if single impurities are unable to do so.  While it is not clear if such currents would occur in real cuprates disordered only by dopant atoms in the high disorder regime, it seems plausible that they might be induced by systematically disordering an overdoped cuprate by electron irradiation, for example.  It is also possible that the physics of highly disordered $d$-wave superconductors, already studied in some detail, may still contain  physics that we do not understand through simple disorder-averaged ``dirty $d$-wave theory"\cite{Hirschfeld1986,SchmittRink1986,Hirschfeld1988,Prohammer1991,felds1993}. Li {\it et al.}\cite{Li2021} showed, for example, that the temperature dependence of the superfluid density in their simulations could be quite linear even in the presence of strong disorder, an unexpected result from the point of view of the disorder-averaged approach.

We stress that the current discussion relates to TRSB arising solely from nonmagnetic disorder in BCS superconductors. It is well-known from a  series of earlier studies that underlying repulsive interactions may cause nonmagnetic disorder to freeze local spin fluctuations, and thereby also break TRS \cite{Tsuchiura2001,ZWang2002,Zhu2002,Chen2004,Andersen2007,Harter2007,Andersen2007,Andersen2010,Schmid_2010,Gastiasoro2013,Gastiasoro2014,Gastiasoro2015,Martiny2015,Gastiasoro2016,Martiny2019}. This mechanism is, however, distinct from the orbital current disorder-induced phases discussed here.  

In this paper, we perform a systematic theoretical study of spontaneous TRSB from nonmagnetic disorder in a $d$-wave superconductor. We analyze the origin of this surprising emergent many-impurity effect, tracing the existence of the currents to the competition between two symmetry-distinct pairing channels. More specifically, we identify the current-carrying regions as areas where sizable extended $s$-wave correlations coexist with the underlying (but disordered) $d$-wave order, making it favorable to locally generate $s\pm id$-like structures. 
This happens despite the fact that the superconducting order in the corresponding homogeneous phase does not support local currents. The phenomenon cannot be understood from a disorder-averaged perspective as a disorder-induced shifting of the $d$ - ($s\pm id$) homogeneous phase boundary; rather it is a local consequence of quantum interference due to particular favorable local impurity configurations. We expect such local orbital current loops to exist quite generically in strongly disordered superconductors exhibiting several competing pairing channels.

\section{Model and Method}

\noindent The study presented here is based on the following one-band BCS mean field Hamiltonian with nearest-neighbor (NN) hopping 
\begin{align}
H & = -\sum_{i,j,\sigma} (t_{ij} + \mu \delta_{ij}) c^{\dagger}_{i,\sigma}c_{j,\sigma} + \sum_{\{p\},\sigma} V_{imp} c^{\dagger}_{p,\sigma}c_{p,\sigma} \nonumber \\
- & \sum_{\langle i,j \rangle} (\Delta_{ij} (c_{i\downarrow}^\dagger c_{j\uparrow}^{\dagger} - c_{i\uparrow}^\dagger c_{j\downarrow}^{\dagger}) + h.c. ) , \label{H}
\end{align}
where $t_{ij} = t$ for NN sites and 0 otherwise, $\Delta_{ij} = V_{0} (\langle c_{j\uparrow} c_{i\downarrow}\rangle - \langle c_{j\downarrow} c_{i\uparrow} \rangle)$ is the NN singlet pairing (where $V_{0} > 0$ ensures attractive pairing) and $\{p\}$ denotes the set of impurity sites. The disorder, which is treated exactly, is incorporated by site-localized scattering potentials, all with strength given by $V_{imp}$. The corresponding Bogoliubov-de Gennes (BdG) equations in the $(c^{\dagger}_{i\uparrow},c_{i\downarrow})$-block of the full Nambu space read
\begin{align}
\sum_{j} \begin{pmatrix} K_{ij} & \Delta_{ij} \\ \Delta^*_{ij} & -K_{ij} \end{pmatrix} \begin{pmatrix} u_{jn\downarrow} \\ v_{jn\uparrow} \end{pmatrix} = E_n \begin{pmatrix} u_{in\downarrow} \\ v_{in\uparrow} \end{pmatrix}, \label{bdg}
\end{align}
where $K_{ij} = -(t_{ij} + \mu\delta_{ij}) + V_{imp} \delta_{ij}\delta_{i \in \{p\}}$ and $u_{in\uparrow}$ ($v_{in\uparrow}$) are the usual coherence factor coefficients from the unitary Bogoliubov transformation. The BdG equations are solved self-consistently on $N \times N$ square lattices with lattice constant $a$, and all results presented below are reported after convergence. The convergence criterion is set to $\sum^{2N^2}_{\alpha, \beta = 1} |H^{BdG}_{\alpha \beta}[m-1] - H^{BdG}_{\alpha \beta}[m]| < 8N^2 \cdot 10^{-10} \, t$ where $m$ indicates iteration number. This criterion ensures that each nonzero input changes no more than $10^{-10}$ on average between two consecutive iterations. All results are computed with periodic boundary conditions, and we fix $t = 1.0$, $V_0=1.5$, and $V_{imp}=3.0$. In addition, we will assume that each impurity dopes one hole into the system, and fix the filling of the system by adjusting the chemical potential, $\mu$, until an average density of $\langle n \rangle = 1 - \rho$, where $\rho$ denotes the dopant fraction, is reached. To calculate the current densities we use that the current on bond $\langle ij \rangle$ is given by
\begin{align}
\langle j_{ij} \rangle &= i\frac{et}{\hbar a^2} \sum_{\sigma} \langle \cd_{i\sigma} c_{j\sigma} - \cd_{j\sigma}c_{i\sigma} \rangle \\
&= -2\frac{et}{\hbar a^2} \sum_n \mathrm{Im} [u^*_{in\downarrow}u_{jn\downarrow} f(E_n) - v^*_{in\uparrow}v_{jn\uparrow}f(E_n)],
\end{align}
where $e$ denotes the electron charge, $a$ is the lattice spacing, and $f(E_n)$ is the Fermi-Dirac distribution function. All results are obtained with the temperature fixed at $T = 0.001 t$. We define the current on a given site $j_i$ as the average of the current on the adjacent bonds, that is
\begin{align}
j^{\hat{x}(\hat{y})}_i = \frac{j_{i,i+\hat{x}(\hat{y})} + j_{i-\hat{x}(\hat{y}),i}}{2}.
\end{align}
The selfconsistency ensures current conservation as confirmed by incoming and outgoing currents being equal in magnitude at all lattice sites.

\section{Results}

\subsection{Homogeneous case}

\begin{figure}[b]
\centering
\includegraphics[angle=0,width=0.99\linewidth]{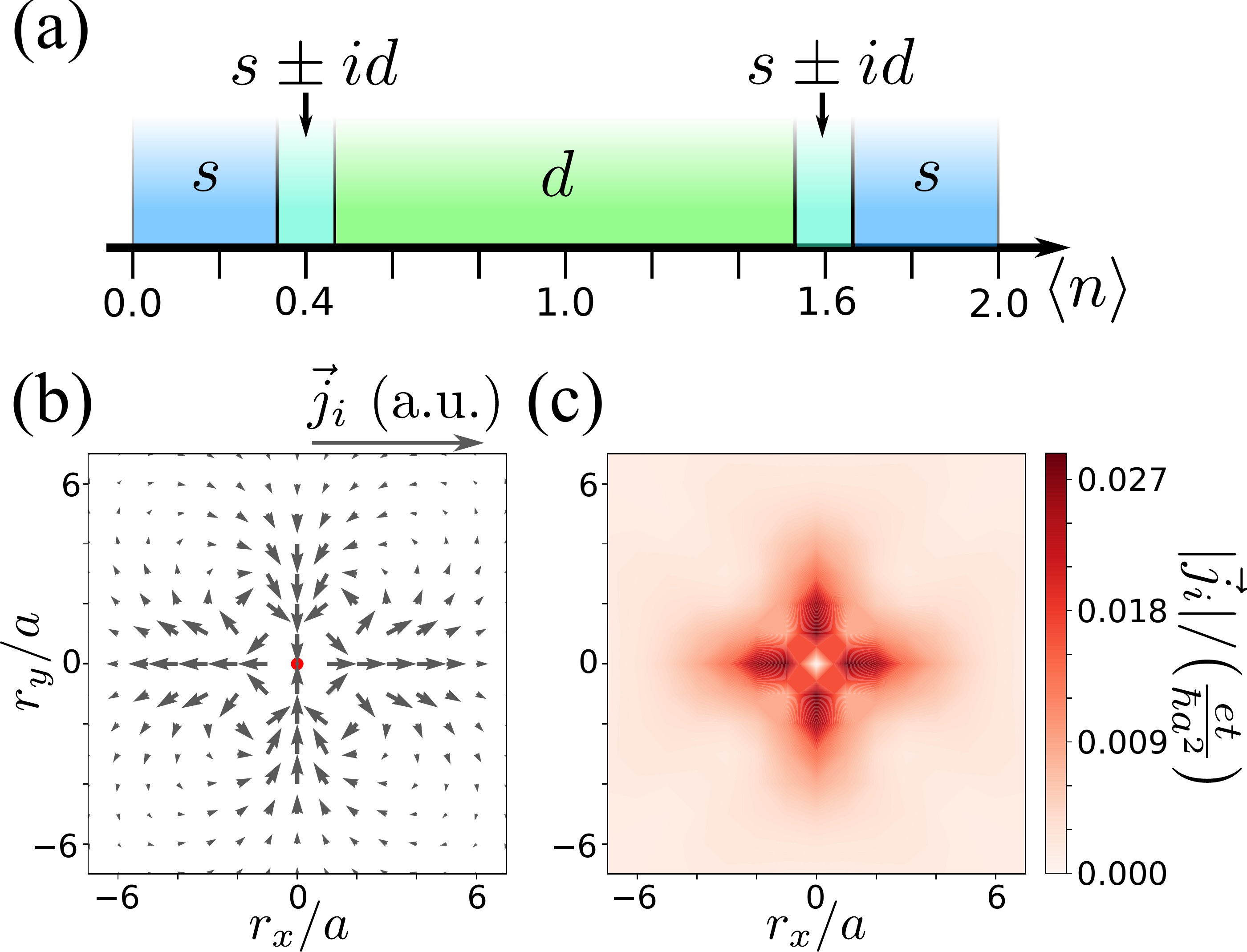}
\caption{(a) Phase diagram of a superconductor with nearest-neighbor attractive pairing as a function of electron density in the homogeneous case. In the crossover regions between $s$- and $d$-wave solutions, the complex $s\pm id$ TRSB superconducting state becomes favorable. (b,c) Impurity-induced currents in the $s+id$ phase at  $\langle n \rangle=0.44$ obtained on a $21\times21$ square lattice. Panel (b) shows the current pattern around a single impurity placed at the center site (red dot), which integrates to zero. All currents with magnitudes $|\vec{j}_i| > 0.25|\vec{j}_i|_{max}$ have been re-scaled to $0.25|\vec{j}_i|_{max}$ for visual clarity. The current is conserved at all sites. (c) Full current amplitude in units of $\frac{et}{\hbar a^2}$. The current magnitudes are interpolated to continuous real-space values for clarity. \label{fig:1}}
\end{figure}

\begin{figure*}[ht]
\centering
\includegraphics[angle=0,width=0.99\linewidth]{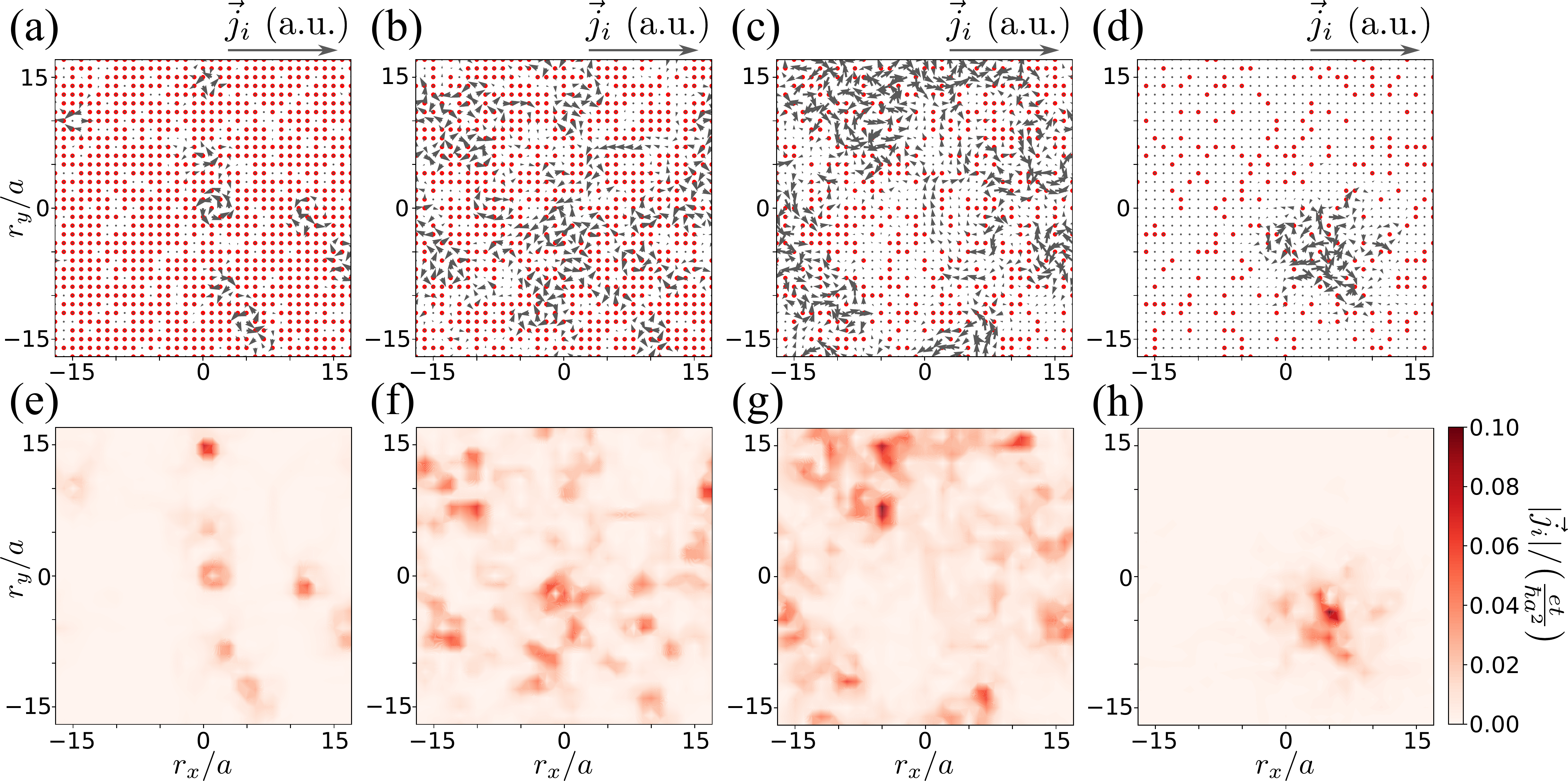}
\caption{Disorder-induced currents for different disorder/doping levels obtained on $35\times35$ square lattices. (a-d) Current patterns for impurity (hole) density $\rho = 0.8$ (a), $0.6$ (b), $0.4$ (c) and $0.2$ (d). Red dots indicate impurity positions. In panels (a-d), all currents with magnitudes $|\vec{j}_i| > 0.25|\vec{j}_i|_{max}$ have been re-scaled to $0.25|\vec{j}_i|_{max}$ for visual clarity. (e-h) Current magnitudes for $\rho = 0.8$ (e), $0.6$ (f), $0.4$ (g) and $0.2$ (h) in units of $\frac{et}{\hbar a^2}$. The current magnitudes are interpolated to continuous real-space values for clarity. \label{fig:2}}
\end{figure*}

\noindent  In the homogeneous system ($\{p\} \in \emptyset$) we find that, depending on the filling, the preferred pairing channel is either extended-$s$, $d$, or $s+id$, see Fig. \ref{fig:1}(a). As we only include attractive NN pairings, $s$ refers to extended NN $s$-wave pairing with nodes along $\cos(k_x) = -\cos(k_y)$, whereas the $d$-wave phase is the $d_{x^2-y^2}$ pairing state with nodes along $\cos(k_x) = \cos(k_y)$. The preferred channel at a given electron density can be understood from the energetics of
gap nodes at the corresponding Fermi surface (FS) \cite{Romer2015,Kreisel2017}. Near half-filling $n=1$, for example, the $s$-wave state is strongly disfavored due to its near alignment of nodes with the FS. By contrast, at very low fillings, the FS consists of a small electron pocket centered at $\Gamma=(0,0)$, which can be only fully gapped by the $s$-wave solution. In the density regimes near $n=0.4, 1.6$, the $s-$ and $d$-wave states become degenerate, and selfconsistent solutions to the gap equation confirm that indeed the preferred low-temperature pairing state is the complex combination $s+id$ (or $s-id$).

\subsection{Single impurity}

While the homogeneous $s\pm id$ state does not support persistent supercurrents, it is well-known that inhomogeneities are able to induce local supercurrents \cite{Lee2009,Garaud2014,Maiti2015,Lin2016}. The simplest situation is a single point-like nonmagnetic impurity with local impurity-induced currents as shown in Fig. \ref{fig:1}(b,c). The amplitude of the impurity-bound currents depends on the impurity strength $V_{imp}$, and saturates for sufficiently large onsite repulsion (or attraction). The $s\pm id$ pairing states are degenerate solutions related through time-reversal. Thus, the current pattern of the $s-id$ pairing state is reversed, but otherwise identical to the result shown in Fig.~\ref{fig:1}(b,c). The emergence of supercurrents in the $s+id$ phase has been previously established mainly through analytical solutions of the associated Ginzburg-Landau equations \cite{Lee2009,Maiti2015}. The result shown in Fig.~\ref{fig:1}(b,c) exhibits a pattern symmetric under $\pi$ rotations, and time-reversal in addition to $\pi/2$-rotation, in agreement with earlier studies \cite{Lee2009,Maiti2015}. As opposed to chiral superconductors, the impurity-induced current from Fig. \ref{fig:1}(b,c) exhibits no net vorticity. The existence of local disorder-induced currents is in stark contrast to pure $d$- or $s$-wave phases where a single nonmagnetic impurity does not generate any supercurrents, and merely features a standard pair-breaking effect. However, as discussed in the next section, impurity-induced spontaneous TRSB does materialize in the presence of a substantial amount of impurities \cite{Li2021}.

\subsection{Multiple impurities}

\begin{figure*}[ht]
\centering
\includegraphics[angle=0,width=0.99\linewidth]{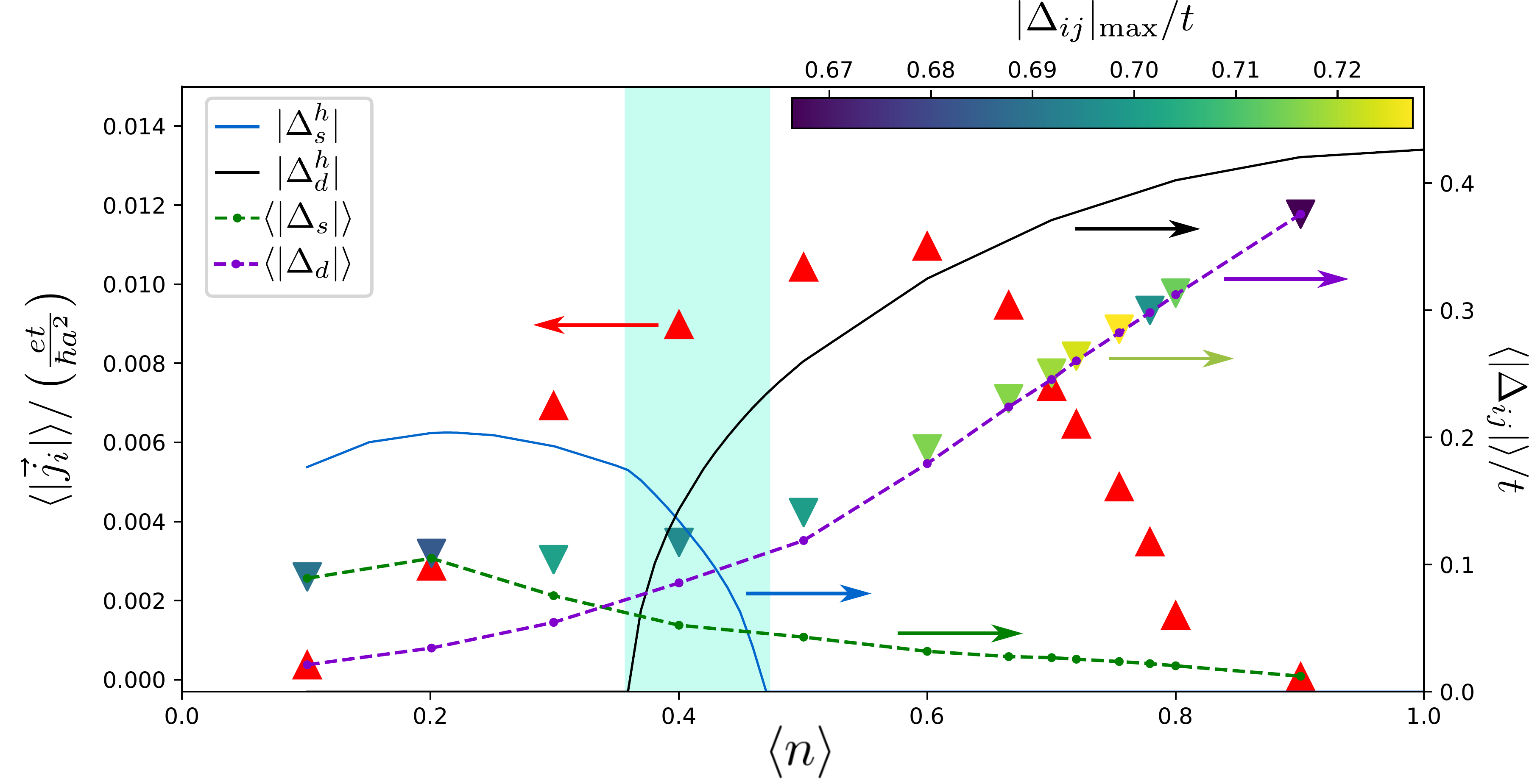}
\caption{Current magnitude, average and maximum bond gap size as a function of average electron density. Red triangles display the total current magnitudes (left $y$-axis), and the positions of inverted triangles show the average gap size (right $y$-axis). Colors of the inverted triangles indicate the maximum bond gap amplitudes. Solid lines are the $s$- and $d$-wave amplitudes ($|\Delta^h_s|$ and $|\Delta^h_d|$, respectively) of the homogeneous system at the corresponding filling. Shaded cyan region marks the homogeneous $s\pm id$ phase as sketched in Fig. \ref{fig:1}(a). Dashed lines show the average $d$-wave (purple) and $s$-wave (green) order parameter components (see main text). The simulations are computed on  $35\times35$ square lattices with random distributions of impurities.\label{fig:3}}
\end{figure*}

\noindent Fig. \ref{fig:2} displays several representative cases of the current density of impurity concentrations $\rho = 80\%, \,60\%, \,40\%$ and $20\%$. The top panels show the direction of the currents at all sites as well as the randomly chosen impurity positions (red dots). The bottom panels display the current magnitudes in units as indicated. The magnitudes are interpolated to continuous real-space values for clarity. By inspection of Fig. \ref{fig:2}, it is evident that substantial currents are generated locally with disorder, producing an emergent current pattern in agreement with Ref.~\onlinecite{Li2021}. This result is rather striking, as only the case in panel (b,f) corresponds to a doping level inside the $s\pm id$ phase region of the clean system. Thus, the nonmagnetic disorder cooperatively breaks TRS locally inside a time-reversal symmetric superconductor. This conclusion is not dependent on the particular value of the impurity potential $V_{imp}$ as long as it is larger than a certain threshold value, whose detailed parameter-dependence we have not further explored.  We stress that the emergent TRSB studied here is a purely orbital effect, distinct from TRSB arising from impurity-induced local magnetic moments caused by nonmagnetic disorder in interacting unconventional superconductors \cite{Tsuchiura2001,ZWang2002,Zhu2002,Chen2004,Andersen2007,Harter2007,Andersen2007,Andersen2010,Schmid_2010,Gastiasoro2013,Gastiasoro2014,Gastiasoro2015,Martiny2015,Gastiasoro2016,Martiny2019}.

In order to quantify the effect, we calculate the impurity configuration-averaged current magnitude, $ \langle |\vec{j}_i| \rangle =N^{-2} \sum_{i} |\vec{j}_i|$, bond gap magnitude, $\langle |\Delta_{ij}| \rangle =\frac{1}{4N^2} \sum_{\langle ij \rangle}|\Delta_{ij}|$, and maximum bond gap amplitude, $|\Delta_{ij}|_{\rm{max}}$, as a function of average electron density $\langle n \rangle = 1 - \rho$, see Fig. \ref{fig:3}. Starting from the clean half-filled case at $\langle n \rangle = 1.0$, we identify a trend of increasing current with increasing doping (i.e. decreasing density away from half-filling) in agreement with the results of Fig.~\ref{fig:2}. Interestingly, the results suggest maximal currents near $\langle n \rangle \simeq 0.6$ rather than at maximal disorder ($\langle n \rangle = 0.5$), or inside the bulk $s\pm i d$ phase. We also identify critical densities of $\sim 0.1, 0.9$ for having sizable currents, yielding an approximately symmetric behaviour around maximal disorder. Regarding the gap amplitudes (inverted triangles in Fig.~\ref{fig:3}), we note a similar trend of $\langle |\Delta_{ij}| \rangle$ as in the homogeneous system (solid lines, $s$-wave amplitude (blue) and $d$-wave amplitude (black)), reflecting the accumulation of states near the normal state Fermi surface as we approach the van Hove singularity at half-filling, $\langle n \rangle = 1.0$. The homogeneous systems consistently exhibit larger average gap amplitudes compared to the impurity-doped systems as expected due to the local disorder-generated gap suppression. We also note that this difference indeed weakens as we approach the undoped/clean limit at $\langle n \rangle = 1.0$. Interestingly, in addition to the general decrease in $\langle |\Delta_{ij}| \rangle$ with doping, we find the lowest $|\Delta_{ij}|_{max}$ (indicated by the color of the inverted triangles in Fig.~\ref{fig:3})  before currents arise ($\langle n \rangle = 0.9$) yielding the least significant gap modulations. Upon further doping, currents are generated and modulations are rapidly enhanced, as is evident from an increase in $|\Delta_{ij}|_{max}$ alongside the decrease in $\langle |\Delta_{ij}| \rangle$. 

\begin{figure*}[ht]
\centering
\includegraphics[angle=0,width=0.99\linewidth]{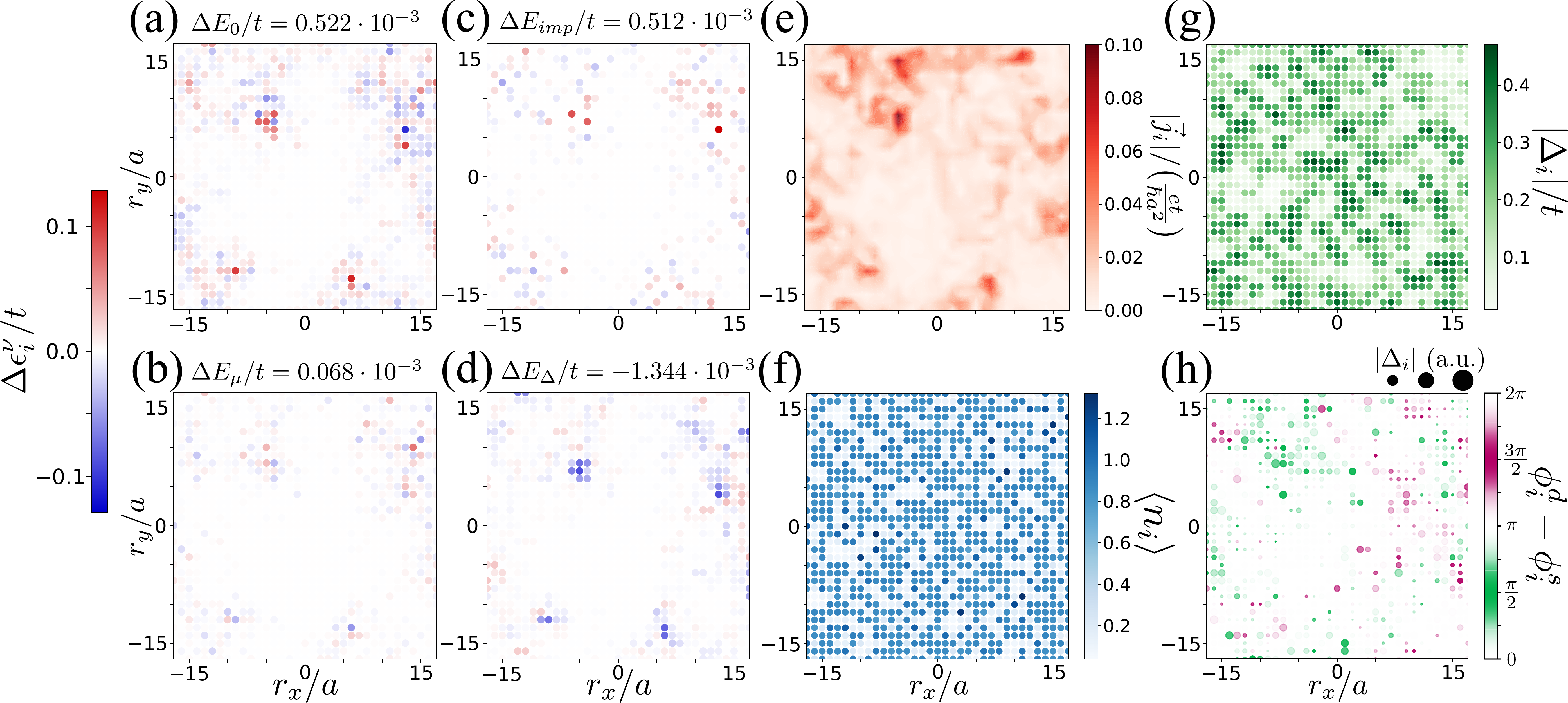}
\caption{(a-d) Difference in free energy density between states breaking and preserving TRS, shown for case (c) of Fig. \ref{fig:2}. The free energy density is divided into contributions, $\epsilon^{0}_i$ (a), $\epsilon^{\mu}_i$ (b), $\epsilon^{imp}_i$ (c), and $\epsilon^{\Delta}_i$ (d), as described in the main text and $\Delta \epsilon^{\nu}_i \equiv \epsilon^{\nu}_{i,j_i \neq 0} - \epsilon^{\nu}_{i,j_i = 0}$ while $\Delta E_{\nu} = N^{-2}\sum_i \Delta \epsilon^{\nu}_i$ is the total energy density difference of the $\nu$'th contribution.  Blue (red) color indicate lowest (highest) free energy density of the TRSB state. The total free energy density gained by breaking TRS is $\Delta E/t = -0.241 \cdot 10^{-3}$. Current magnitude (e), local density (f) and site-centered gap amplitude, $|\Delta_i|=\frac{1}{4}\sum_\delta |\Delta_{i,i+\delta}|$, (g) of a representative $35\times35$ system with a TRSB ground state. Current pattern and impurity configuration are shown in Fig.~\ref{fig:2}(c). (h) Local phase difference of local $s$- and $d$-wave order parameters (color scale) and site-centered gap amplitude (symbol size).  \label{fig:4}}
\end{figure*}

We define the site-centered $s$-wave ($d$-wave) component by the standard expressions
\begin{align}
\Delta_i^{s(d)} = \frac{\Delta_{i,i+\hat{x}} + \Delta_{i-\hat{x},i} +(-) \hat{x}\leftrightarrow \hat{y}}{4}, \label{sdorder}
\end{align}
and thereby obtain the average $s$-wave ($d$-wave) contribution, $\langle |\Delta_{s(d)}| \rangle = N^{-2} \sum_i |\Delta_i^{s(d)}|$, shown by the dashed green (purple) line in Fig.~\ref{fig:3}. As seen, there is a substantial $s$-wave component induced even in the range $\langle n \rangle \simeq 0.45-0.9$, where $d$-wave pairing is the ground state in the corresponding clean system. This is a multiple-impurity effect, setting in as disorder sites become dense enough and break the point group symmetries. Thus, a consequence of introducing large enough disorder levels is that inhomogeneous puddles of extended $s$-wave ($d$-wave) order leaks into the $d$-wave ($s$-wave) regions of the clean system.

Figure~\ref{fig:3} strongly suggests that the emergence of currents in a disordered $d$-wave superconductor of the form studied here is generally energetically favorable. To elucidate the lowering of energy associated with the generation of currents, we compute the low-temperature free energy densities of identical systems with and without currents. We converge the state without currents by choosing (random) real initial values of all gaps (as opposed to complex initial values in all other computations). Importantly, the computation of the free energies indeed confirm that the phases {\it with} currents constitute the ground state. Next, we separate the different contributions of the total energy into the following parts
\begin{eqnarray}
\epsilon^0_i &=&  -\frac{t}{2}\sum_{\delta, \sigma} \langle c^{\dagger}_{i,\sigma}c_{i+\delta,\sigma} \rangle,  \label{eq:eps_i_0}\\
\epsilon^{\mu}_i &=&  -\mu \sum_{\sigma} \langle n_{i,\sigma} \rangle, \\
\epsilon^{\Delta}_i &=& -  \sum_{\delta} \frac{|\Delta_{i,i+\delta}|^2}{2V_0}, \label{eq:delta}\\
\epsilon^{imp}_i &=& \sum_{\sigma} V_{imp} \delta_{i \in p} \langle n_{i,\sigma} \rangle,\label{eq:eps_i_imp}
\end{eqnarray}
where $\delta = \pm \hat{x},\pm \hat{y}$. We further define $E_{\nu} \equiv \frac{1}{N^2} \sum_i \epsilon_i^{\nu}$, where $\epsilon_i^\nu$ is the $\nu$th contribution (\ref{eq:eps_i_0})-(\ref{eq:eps_i_imp}), such that $\sum_\nu E_\nu$ is the total energy.  In Fig.~\ref{fig:4}(a-d), we show a representative example of the energy density difference between the state with and the state without induced currents, i.e. $\Delta \epsilon^{\nu}_i = \epsilon^{\nu}_{i,j_i \neq 0} - \epsilon^{\nu}_{i,j_i = 0}$. Thus, for negative (positive) values, the state with currents has lower (higher) energy density compared to the state without currents. The results shown in Fig.~\ref{fig:4} correspond to an impurity configuration with $\rho = 40 \%$ equivalent to $\langle n \rangle = 0.6$. The results suggest that  generating currents leads to an overall pairing energy gain. Averaging over ten different impurity configurations confirms this result. Thus, the current formation is driven by an energy gain in the pairing term, $\Delta E_{\Delta}/t$. (We find this result to be generally valid beyond approximately  $\rho \simeq 0.25$. Below this disorder level, i.e. closer to the clean half-filled case, the energy gain from current-carrying ground states also acquires significant contributions from density distributions, i.e. $\Delta E_{imp}$ and $\Delta E_{\mu}$.)

From Fig.~\ref{fig:4}(d) and \ref{fig:4}(e), one can visually detect a correlation between the local gain in pairing energy and the local currents. Additionally, one sees that the currents are not well correlated with either the electron density (Fig.~\ref{fig:4}(f)) or the total gap amplitudes (Fig.~\ref{fig:4}(g)). Rather, typical areas of substantial local current amplitudes appear well correlated with a $\pi/2$ or $3\pi/2$ phase difference $\phi_i^d-\phi_i^s$ between the corresponding $s$- and $d$-wave order parameters, Eq.~({\ref{sdorder}}).  To quantify these comparisons, the correlation coefficients are obtained by 
\begin{align}
corr\{x,y\}= \frac{\langle xy \rangle - \langle x \rangle \langle y \rangle}{\sqrt{\langle x^2 \rangle - \langle x \rangle^2}\sqrt{\langle y^2 \rangle - \langle y \rangle^2}},
\end{align}
for nine different impurity configurations, resulting in configuration-averaged values of $corr\{|\vec{j}_i|, \langle n_i \rangle \} = 0.193$, $corr\{|\vec{j}_i|,|\Delta_i|\} = 0.184$ and $corr\{|\vec{j}_i|,|\rm{Im}[e^{i(\phi^d_i - \phi^s_i)}]| |\Delta_i|\} = 0.334$. The  latter reflects the two-fold requirement of a substantial local pairing amplitude ($|\Delta_i|$) combined with a phase difference of $\phi^d_i - \phi^s_i \sim \pi/2, \,3\pi/2$ (yielding large values of $|\rm{Im}[e^{i(\phi^d_i - \phi^s_i)}]|$). It should be noted that the correlations generally increase for doping levels further away from maximal disorder ($\langle n \rangle = 0.5$), and for the result shown in Fig.~\ref{fig:2}(a), for example, $corr\{|\vec{j}_i|,|\rm{Im}[e^{i(\phi^d_i - \phi^s_i)}]| |\Delta_i|\} = 0.625$ while $corr\{|\vec{j}_i|, \langle n_i \rangle \} = 0.234$ and $corr\{|\vec{j}_i|,|\Delta_i|\} = 0.262$. Thus, the reason for the induced local currents is the following: the presence of significant disorder leads to strong inhomogeneity in the gap suppression. This inevitably induces spatially varying $s$-wave ($d$-wave) order in the corresponding clean $d$-wave ($s$-wave) phase. In certain regions it becomes energetically favorable to generate local $s\pm id$ structures, where currents arise due to density modulations. This appears similar
to the single-impurity results shown in Fig.~\ref{fig:1}, except that the complex mixing of the two components occurs only locally, in a doping range where the homogeneous $s+id$ phase is unstable. We note that the correlation between currents and the phase difference $\phi_i^d-\phi_i^s$ is not perfect mainly since some local regions of $\phi_i^d-\phi_i^s=\pi/2,\,3\pi/2$ do not support currents. This can be understood from the fact that in an $s\pm id$ superconductor, currents are generated not only by phase gradients, but also by spatial modulations of the pairing amplitude \cite{Lee2009}. In some regions the local phase and amplitude modulations of $\Delta_{ij}$ "destructively interfere" and lead to vanishing currents. As such, the combination of a sizable local gap amplitude and $s \pm id$ pairing is a necessary but not sufficient condition for the emergence of currents in the surrounding area.

\section{Discussion and conclusions}

\noindent We now discuss several  different plausible explanations for the mechanism behind the disorder-induced currents. 
We have already shown that the current-carrying state is the ground state, so that these solutions do not correspond to metastable free energy minima in a complex disorder potential landscape.  
We further consider meta-stability arising from the neglect of proper feedback/selfconsistency of the induced currents and their generated vector potential. To investigate this, we include the Peierls phases by solving the Maxwell equations and let $t \rightarrow t \exp[i\frac{e}{\hbar} \int \! \mathbf{A}\cdot d\mathbf{l}]$ in the selfconsistent solution. The small resulting (dimensionless) vector potential $ \frac{ea}{\hbar} \mathbf{A} = \mathbf{\tilde{A}} \sim \mathcal{O}(10^{-6})$, yields essentially no change in the preferred ground state. Therefore, we conclude that inclusion of the vector potential generated by the currents leads to only minor quantitative changes to the above results. 

We have also explored the idea of local doping into an $s\pm id$ phase, i.e. whether the currents can be understood as areas where the average densities are similar to the bulk $s\pm id$ phase near $\langle n \rangle\simeq 0.4$. This is clearly not the case, however, as the disorder is much too strong for such "local density approximations", and furthermore the electron density is only very weakly correlated with the current maps. 

\begin{figure}[b]
\centering
\includegraphics[angle=0,width=0.99\linewidth]{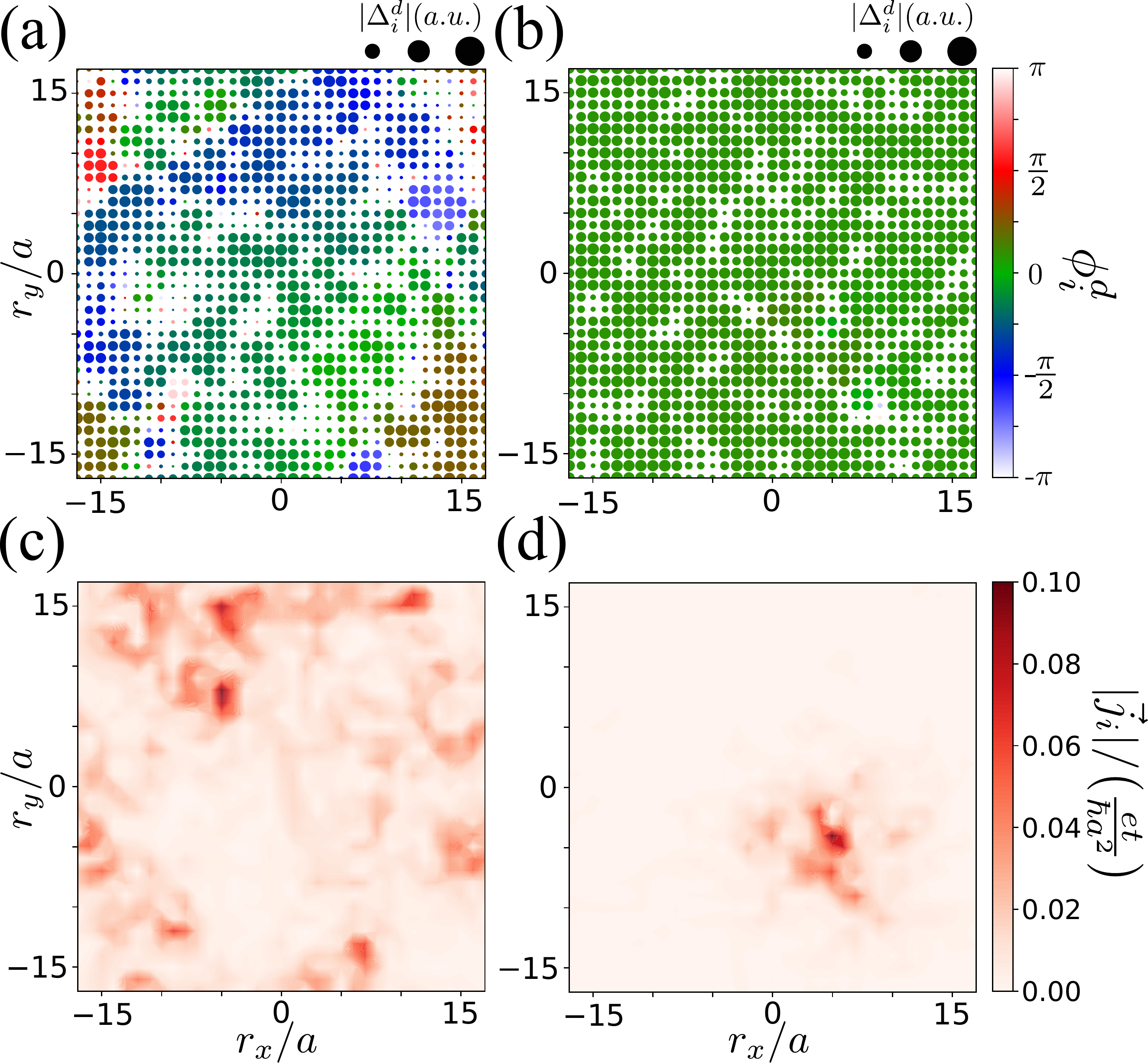}
\caption{(a,b) Local phase, $\phi^i_d$ (color scale), and amplitude, $|\Delta^i_d|$ (symbol size),
of the $d$-wave component for the cases in panels (c) and (d) of Fig.~\ref{fig:2}, respectively. (c,d) Same as Fig.~\ref{fig:2}(g,h) included here for visual comparison.\label{fig:5}}
\end{figure}

Finally, we have investigated to what extent the local currents arise from disorder in the phase of the $d$-wave component \cite{Li2021,Phase_impurities}. As evident from Fig.~\ref{fig:3}, this component is dominant for all $\langle n \rangle \gtrsim 0.5$ and, as such, it is not unreasonable to assume that the generation of currents arises mainly from modulations of this component. In Fig.~\ref{fig:5}(a) we show the local $d$-wave phase, $\phi^i_d$, and amplitude, $|\Delta^i_d|$, represented by the color scale and symbol size, respectively. The converged result presented here is the same as in Fig.~\ref{fig:4} and, by comparing to Fig.~\ref{fig:4}(g) and Fig.~\ref{fig:5}(a), it is clear that the $d$-wave component does indeed capture the essential spatial variations of the local pairing amplitudes.
Strong amplitude modulations are evident in Fig.~\ref{fig:5}(a), and some grain formation in the $d$-wave phase is apparent. By visual comparison of Fig.~\ref{fig:5}(a) and Fig.~\ref{fig:5}(c), several domain walls do indeed coincide with current-carrying regions. This is to be expected, as phase gradients should give rise to currents. However, computing the configuration-averaged correlation for nine different impurity distributions at $\langle n \rangle = 0.6$ yields $corr\{|\vec{j}_i|,|\Delta^i_d||\nabla \phi^i_d|\} = 0.196$, significantly lower than the correlation between current magnitudes and regions of local $s\pm id$ pairing. Additionally, in Fig.~\ref{fig:5}(b,d) we show the same quantities for a system with $\langle n \rangle = 0.8$. Here phase modulations are barely visible and no grain formation has taken place despite the presence of currents. Thus, we conclude that emergent weakly Josephson coupled $d$-wave grains are not the main mechanism leading to the formation of currents.

In summary, we have performed a theoretical study of nonmagnetic disorder-induced orbital current loops within a model where superconductivity is stabilized by nearest-neighbor attraction. In the homogeneous case, this model supports both extended $s$-wave and $d$-wave superconductivity, depending on doping level. At the degeneracy point of these two orders, the complex combinations $s\pm i d$ are preferred. However, in the disordered case, even far away from the doping regime where homogeneous $s\pm i d$ order exists, certain favorable regions support local $s\pm i d$ structures, and induce currents. Correspondingly, models supporting only a single pairing channel do not lead to disorder-induced orbital currents of the form discussed here~\cite{Li2021}. In terms of experiments, based on the explanation provided above we expect TRSB from orbital currents in any disordered superconductors with several competing pairing instabilities, as potentially relevant for e.g Sr$_2$RuO$_4$ \cite{Luke1998}, UTe$_2$, and Ba$_{1-x}$K$_x$Fe$_2$As$_2$ \cite{Grinenko2020}. It remains to be further explored what are the unique experimental litmus tests of the local currents, but clearly $\mu$SR may pick up the fields generated by the loop currents below $T_c$. As shown here, it is not necessary for the clean systems to reside in a TRSB phase; disorder alone can induce local TRSB regions supporting current loops. It remains an interesting outstanding theoretical question how disorder-induced currents are affected by both thermal and quantum fluctuations.

\section*{Acknowledgements} We acknowledge fruitful discussions with S. A. Kivelson and A. T. R\o mer. C. N. B. and B. M. A. acknowledge support by the Danish national Committee for Research Infrastructure (NUFI) through the ESS-Lighthouse Q-MAT. B. M. A. acknowledges support from the Independent Research Fund Denmark grant number 8021-00047B. P.J.H. acknowledges support from NSF-DMR-1849751.

 \bibliography{litlist}

\begin{thebibliography}{42}%
\makeatletter
\providecommand \@ifxundefined [1]{%
 \@ifx{#1\undefined}
}%
\providecommand \@ifnum [1]{%
 \ifnum #1\expandafter \@firstoftwo
 \else \expandafter \@secondoftwo
 \fi
}%
\providecommand \@ifx [1]{%
 \ifx #1\expandafter \@firstoftwo
 \else \expandafter \@secondoftwo
 \fi
}%
\providecommand \natexlab [1]{#1}%
\providecommand \enquote  [1]{``#1''}%
\providecommand \bibnamefont  [1]{#1}%
\providecommand \bibfnamefont [1]{#1}%
\providecommand \citenamefont [1]{#1}%
\providecommand \href@noop [0]{\@secondoftwo}%
\providecommand \href [0]{\begingroup \@sanitize@url \@href}%
\providecommand \@href[1]{\@@startlink{#1}\@@href}%
\providecommand \@@href[1]{\endgroup#1\@@endlink}%
\providecommand \@sanitize@url [0]{\catcode `\\12\catcode `\$12\catcode
  `\&12\catcode `\#12\catcode `\^12\catcode `\_12\catcode `\%12\relax}%
\providecommand \@@startlink[1]{}%
\providecommand \@@endlink[0]{}%
\providecommand \url  [0]{\begingroup\@sanitize@url \@url }%
\providecommand \@url [1]{\endgroup\@href {#1}{\urlprefix }}%
\providecommand \urlprefix  [0]{URL }%
\providecommand \Eprint [0]{\href }%
\providecommand \doibase [0]{https://doi.org/}%
\providecommand \selectlanguage [0]{\@gobble}%
\providecommand \bibinfo  [0]{\@secondoftwo}%
\providecommand \bibfield  [0]{\@secondoftwo}%
\providecommand \translation [1]{[#1]}%
\providecommand \BibitemOpen [0]{}%
\providecommand \bibitemStop [0]{}%
\providecommand \bibitemNoStop [0]{.\EOS\space}%
\providecommand \EOS [0]{\spacefactor3000\relax}%
\providecommand \BibitemShut  [1]{\csname bibitem#1\endcsname}%
\let\auto@bib@innerbib\@empty
\bibitem [{\citenamefont {Vollhardt}\ and\ \citenamefont
  {W{\"o}lfle}(1990)}]{VollhardtWoelfle}%
  \BibitemOpen
  \bibfield  {author} {\bibinfo {author} {\bibfnamefont {D.}~\bibnamefont
  {Vollhardt}}\ and\ \bibinfo {author} {\bibfnamefont {P.}~\bibnamefont
  {W{\"o}lfle}},\ }\href@noop {} {\emph {\bibinfo {title} {The Superfluid
  Phases of Helium-3}}},\ \bibinfo {edition} {1st}\ ed.\ (\bibinfo  {publisher}
  {Taylor and Francis},\ \bibinfo {address} {New York},\ \bibinfo {year}
  {1990})\BibitemShut {NoStop}%
\bibitem [{\citenamefont {Mackenzie}\ \emph {et~al.}(2017)\citenamefont
  {Mackenzie}, \citenamefont {Scaffidi}, \citenamefont {Hicks},\ and\
  \citenamefont {Maeno}}]{Mackenzie2017}%
  \BibitemOpen
  \bibfield  {author} {\bibinfo {author} {\bibfnamefont {A.~P.}\ \bibnamefont
  {Mackenzie}}, \bibinfo {author} {\bibfnamefont {T.}~\bibnamefont {Scaffidi}},
  \bibinfo {author} {\bibfnamefont {C.~W.}\ \bibnamefont {Hicks}},\ and\
  \bibinfo {author} {\bibfnamefont {Y.}~\bibnamefont {Maeno}},\ }\bibfield
  {title} {\bibinfo {title} {Even odder after twenty-three years: the
  superconducting order parameter puzzle of sr2ruo4},\ }\href
  {https://doi.org/10.1038/s41535-017-0045-4} {\bibfield  {journal} {\bibinfo
  {journal} {npj Quantum Materials}\ }\textbf {\bibinfo {volume} {2}},\
  \bibinfo {pages} {40} (\bibinfo {year} {2017})}\BibitemShut {NoStop}%
\bibitem [{\citenamefont {Sauls}(1994)}]{Sauls2018}%
  \BibitemOpen
  \bibfield  {author} {\bibinfo {author} {\bibfnamefont {J.}~\bibnamefont
  {Sauls}},\ }\bibfield  {title} {\bibinfo {title} {The order parameter for the
  superconducting phases of upt3},\ }\href
  {https://doi.org/10.1080/00018739400101475} {\bibfield  {journal} {\bibinfo
  {journal} {Advances in Physics}\ }\textbf {\bibinfo {volume} {43}},\ \bibinfo
  {pages} {113} (\bibinfo {year} {1994})}\BibitemShut {NoStop}%
\bibitem [{\citenamefont {Moore}\ and\ \citenamefont
  {Read}(1991)}]{MooreRead1991}%
  \BibitemOpen
  \bibfield  {author} {\bibinfo {author} {\bibfnamefont {G.}~\bibnamefont
  {Moore}}\ and\ \bibinfo {author} {\bibfnamefont {N.}~\bibnamefont {Read}},\
  }\bibfield  {title} {\bibinfo {title} {Nonabelions in the fractional quantum
  hall effect},\ }\href
  {https://doi.org/https://doi.org/10.1016/0550-3213(91)90407-O} {\bibfield
  {journal} {\bibinfo  {journal} {Nuclear Physics B}\ }\textbf {\bibinfo
  {volume} {360}},\ \bibinfo {pages} {362} (\bibinfo {year}
  {1991})}\BibitemShut {NoStop}%
\bibitem [{\citenamefont {Pustogow}\ \emph {et~al.}(2019)\citenamefont
  {Pustogow}, \citenamefont {Luo}, \citenamefont {Chronister}, \citenamefont
  {Su}, \citenamefont {Sokolov}, \citenamefont {Jerzembeck}, \citenamefont
  {Mackenzie}, \citenamefont {Hicks}, \citenamefont {Kikugawa}, \citenamefont
  {Raghu}, \citenamefont {Bauer},\ and\ \citenamefont {Brown}}]{Pustogow19}%
  \BibitemOpen
  \bibfield  {author} {\bibinfo {author} {\bibfnamefont {A.}~\bibnamefont
  {Pustogow}}, \bibinfo {author} {\bibfnamefont {Y.}~\bibnamefont {Luo}},
  \bibinfo {author} {\bibfnamefont {A.}~\bibnamefont {Chronister}}, \bibinfo
  {author} {\bibfnamefont {Y.-S.}\ \bibnamefont {Su}}, \bibinfo {author}
  {\bibfnamefont {D.~A.}\ \bibnamefont {Sokolov}}, \bibinfo {author}
  {\bibfnamefont {F.}~\bibnamefont {Jerzembeck}}, \bibinfo {author}
  {\bibfnamefont {A.~P.}\ \bibnamefont {Mackenzie}}, \bibinfo {author}
  {\bibfnamefont {C.~W.}\ \bibnamefont {Hicks}}, \bibinfo {author}
  {\bibfnamefont {N.}~\bibnamefont {Kikugawa}}, \bibinfo {author}
  {\bibfnamefont {S.}~\bibnamefont {Raghu}}, \bibinfo {author} {\bibfnamefont
  {E.~D.}\ \bibnamefont {Bauer}},\ and\ \bibinfo {author} {\bibfnamefont
  {S.~E.}\ \bibnamefont {Brown}},\ }\bibfield  {title} {\bibinfo {title}
  {Constraints on the superconducting order parameter in sr$_2$ruo$_4$ from
  oxygen-17 nuclear magnetic resonance},\ }\href
  {https://doi.org/10.1038/s41586-019-1596-2} {\bibfield  {journal} {\bibinfo
  {journal} {Nature}\ }\textbf {\bibinfo {volume} {574}},\ \bibinfo {pages}
  {72} (\bibinfo {year} {2019})}\BibitemShut {NoStop}%
\bibitem [{\citenamefont {Chronister}\ \emph {et~al.}(2021)\citenamefont
  {Chronister}, \citenamefont {Pustogow}, \citenamefont {Kikugawa},
  \citenamefont {Sokolov}, \citenamefont {Jerzembeck}, \citenamefont {Hicks},
  \citenamefont {Mackenzie}, \citenamefont {Bauer},\ and\ \citenamefont
  {Brown}}]{Chronister}%
  \BibitemOpen
  \bibfield  {author} {\bibinfo {author} {\bibfnamefont {A.}~\bibnamefont
  {Chronister}}, \bibinfo {author} {\bibfnamefont {A.}~\bibnamefont
  {Pustogow}}, \bibinfo {author} {\bibfnamefont {N.}~\bibnamefont {Kikugawa}},
  \bibinfo {author} {\bibfnamefont {D.~A.}\ \bibnamefont {Sokolov}}, \bibinfo
  {author} {\bibfnamefont {F.}~\bibnamefont {Jerzembeck}}, \bibinfo {author}
  {\bibfnamefont {C.~W.}\ \bibnamefont {Hicks}}, \bibinfo {author}
  {\bibfnamefont {A.~P.}\ \bibnamefont {Mackenzie}}, \bibinfo {author}
  {\bibfnamefont {E.~D.}\ \bibnamefont {Bauer}},\ and\ \bibinfo {author}
  {\bibfnamefont {S.~E.}\ \bibnamefont {Brown}},\ }\bibfield  {title} {\bibinfo
  {title} {Evidence for even parity unconventional superconductivity in
  {Sr}$_2${RuO}$_4$},\ }\href {https://www.pnas.org/content/118/25/e2025313118}
  {\bibfield  {journal} {\bibinfo  {journal} {Proceedings of the National
  Academy of Sciences}\ }\textbf {\bibinfo {volume} {118}} (\bibinfo {year}
  {2021})}\BibitemShut {NoStop}%
\bibitem [{\citenamefont {Luke}\ \emph {et~al.}(1998)\citenamefont {Luke},
  \citenamefont {Fudamoto}, \citenamefont {Kojima}, \citenamefont {Larkin},
  \citenamefont {Merrin}, \citenamefont {Nachumi}, \citenamefont {Uemura},
  \citenamefont {Maeno}, \citenamefont {Mao}, \citenamefont {Mori},
  \citenamefont {Nakamura},\ and\ \citenamefont {Sigrist}}]{Luke1998}%
  \BibitemOpen
  \bibfield  {author} {\bibinfo {author} {\bibfnamefont {G.~M.}\ \bibnamefont
  {Luke}}, \bibinfo {author} {\bibfnamefont {Y.}~\bibnamefont {Fudamoto}},
  \bibinfo {author} {\bibfnamefont {K.~M.}\ \bibnamefont {Kojima}}, \bibinfo
  {author} {\bibfnamefont {M.~I.}\ \bibnamefont {Larkin}}, \bibinfo {author}
  {\bibfnamefont {J.}~\bibnamefont {Merrin}}, \bibinfo {author} {\bibfnamefont
  {B.}~\bibnamefont {Nachumi}}, \bibinfo {author} {\bibfnamefont {Y.~J.}\
  \bibnamefont {Uemura}}, \bibinfo {author} {\bibfnamefont {Y.}~\bibnamefont
  {Maeno}}, \bibinfo {author} {\bibfnamefont {Z.~Q.}\ \bibnamefont {Mao}},
  \bibinfo {author} {\bibfnamefont {Y.}~\bibnamefont {Mori}}, \bibinfo {author}
  {\bibfnamefont {H.}~\bibnamefont {Nakamura}},\ and\ \bibinfo {author}
  {\bibfnamefont {M.}~\bibnamefont {Sigrist}},\ }\bibfield  {title} {\bibinfo
  {title} {Time-reversal symmetry-breaking superconductivity in \sruo},\ }\href
  {https://doi.org/10.1038/29038} {\bibfield  {journal} {\bibinfo  {journal}
  {Nature}\ }\textbf {\bibinfo {volume} {394}},\ \bibinfo {pages} {558}
  (\bibinfo {year} {1998})}\BibitemShut {NoStop}%
\bibitem [{\citenamefont {Kapitulnik}\ \emph {et~al.}(2009)\citenamefont
  {Kapitulnik}, \citenamefont {Xia}, \citenamefont {Schemm},\ and\
  \citenamefont {Palevski}}]{Kapitulnik09}%
  \BibitemOpen
  \bibfield  {author} {\bibinfo {author} {\bibfnamefont {A.}~\bibnamefont
  {Kapitulnik}}, \bibinfo {author} {\bibfnamefont {J.}~\bibnamefont {Xia}},
  \bibinfo {author} {\bibfnamefont {E.}~\bibnamefont {Schemm}},\ and\ \bibinfo
  {author} {\bibfnamefont {A.}~\bibnamefont {Palevski}},\ }\bibfield  {title}
  {\bibinfo {title} {Polar {K}err effect as probe for time-reversal symmetry
  breaking in unconventional superconductors},\ }\href
  {https://doi.org/10.1088/1367-2630/11/5/055060} {\bibfield  {journal}
  {\bibinfo  {journal} {New Journal of Physics}\ }\textbf {\bibinfo {volume}
  {11}},\ \bibinfo {pages} {055060} (\bibinfo {year} {2009})}\BibitemShut
  {NoStop}%
\bibitem [{\citenamefont {R\o{}mer}\ \emph {et~al.}(2019)\citenamefont
  {R\o{}mer}, \citenamefont {Scherer}, \citenamefont {Eremin}, \citenamefont
  {Hirschfeld},\ and\ \citenamefont {Andersen}}]{RomerPRL}%
  \BibitemOpen
  \bibfield  {author} {\bibinfo {author} {\bibfnamefont {A.~T.}\ \bibnamefont
  {R\o{}mer}}, \bibinfo {author} {\bibfnamefont {D.~D.}\ \bibnamefont
  {Scherer}}, \bibinfo {author} {\bibfnamefont {I.~M.}\ \bibnamefont {Eremin}},
  \bibinfo {author} {\bibfnamefont {P.~J.}\ \bibnamefont {Hirschfeld}},\ and\
  \bibinfo {author} {\bibfnamefont {B.~M.}\ \bibnamefont {Andersen}},\
  }\bibfield  {title} {\bibinfo {title} {Knight shift and leading
  superconducting instability from spin fluctuations in \sruo},\ }\href
  {https://doi.org/10.1103/PhysRevLett.123.247001} {\bibfield  {journal}
  {\bibinfo  {journal} {Phys. Rev. Lett.}\ }\textbf {\bibinfo {volume} {123}},\
  \bibinfo {pages} {247001} (\bibinfo {year} {2019})}\BibitemShut {NoStop}%
\bibitem [{\citenamefont {Rømer}\ and\ \citenamefont
  {Andersen}(2020)}]{Romer_MPLB}%
  \BibitemOpen
  \bibfield  {author} {\bibinfo {author} {\bibfnamefont {A.~T.}\ \bibnamefont
  {Rømer}}\ and\ \bibinfo {author} {\bibfnamefont {B.~M.}\ \bibnamefont
  {Andersen}},\ }\bibfield  {title} {\bibinfo {title} {Fluctuation-driven
  superconductivity in sr2ruo4 from weak repulsive interactions},\ }\href
  {https://doi.org/10.1142/S0217984920400527} {\bibfield  {journal} {\bibinfo
  {journal} {Modern Physics Letters B}\ }\textbf {\bibinfo {volume} {34}},\
  \bibinfo {pages} {2040052} (\bibinfo {year} {2020})}\BibitemShut {NoStop}%
\bibitem [{\citenamefont {R\o{}mer}\ \emph {et~al.}(2021)\citenamefont
  {R\o{}mer}, \citenamefont {Hirschfeld},\ and\ \citenamefont
  {Andersen}}]{Romer2021_Longerrange}%
  \BibitemOpen
  \bibfield  {author} {\bibinfo {author} {\bibfnamefont {A.~T.}\ \bibnamefont
  {R\o{}mer}}, \bibinfo {author} {\bibfnamefont {P.~J.}\ \bibnamefont
  {Hirschfeld}},\ and\ \bibinfo {author} {\bibfnamefont {B.~M.}\ \bibnamefont
  {Andersen}},\ }\bibfield  {title} {\bibinfo {title} {Superconducting state of
  ${\mathrm{sr}}_{2}{\mathrm{ruo}}_{4}$ in the presence of longer-range coulomb
  interactions},\ }\href {https://doi.org/10.1103/PhysRevB.104.064507}
  {\bibfield  {journal} {\bibinfo  {journal} {Phys. Rev. B}\ }\textbf {\bibinfo
  {volume} {104}},\ \bibinfo {pages} {064507} (\bibinfo {year}
  {2021})}\BibitemShut {NoStop}%
\bibitem [{\citenamefont {Clepkens}\ \emph {et~al.}(2021)\citenamefont
  {Clepkens}, \citenamefont {Lindquist},\ and\ \citenamefont
  {Kee}}]{clepkens2021}%
  \BibitemOpen
  \bibfield  {author} {\bibinfo {author} {\bibfnamefont {J.}~\bibnamefont
  {Clepkens}}, \bibinfo {author} {\bibfnamefont {A.~W.}\ \bibnamefont
  {Lindquist}},\ and\ \bibinfo {author} {\bibfnamefont {H.-Y.}\ \bibnamefont
  {Kee}},\ }\bibfield  {title} {\bibinfo {title} {Shadowed triplet pairings in
  hund's metals with spin-orbit coupling},\ }\href
  {https://doi.org/10.1103/PhysRevResearch.3.013001} {\bibfield  {journal}
  {\bibinfo  {journal} {Phys. Rev. Research}\ }\textbf {\bibinfo {volume}
  {3}},\ \bibinfo {pages} {013001} (\bibinfo {year} {2021})}\BibitemShut
  {NoStop}%
\bibitem [{\citenamefont {Kivelson}\ \emph {et~al.}(2020)\citenamefont
  {Kivelson}, \citenamefont {Yuan}, \citenamefont {Ramshaw},\ and\
  \citenamefont {Thomale}}]{kivelson2020proposal}%
  \BibitemOpen
  \bibfield  {author} {\bibinfo {author} {\bibfnamefont {S.~A.}\ \bibnamefont
  {Kivelson}}, \bibinfo {author} {\bibfnamefont {A.~C.}\ \bibnamefont {Yuan}},
  \bibinfo {author} {\bibfnamefont {B.}~\bibnamefont {Ramshaw}},\ and\ \bibinfo
  {author} {\bibfnamefont {R.}~\bibnamefont {Thomale}},\ }\bibfield  {title}
  {\bibinfo {title} {A proposal for reconciling diverse experiments on the
  superconducting state in sr2ruo4},\ }\href@noop {} {\bibfield  {journal}
  {\bibinfo  {journal} {npj Quantum Materials}\ }\textbf {\bibinfo {volume}
  {5}},\ \bibinfo {pages} {43} (\bibinfo {year} {2020})}\BibitemShut {NoStop}%
\bibitem [{\citenamefont {Karmakar}\ and\ \citenamefont
  {Ganesh}(2021)}]{Karmakar2021}%
  \BibitemOpen
  \bibfield  {author} {\bibinfo {author} {\bibfnamefont {M.}~\bibnamefont
  {Karmakar}}\ and\ \bibinfo {author} {\bibfnamefont {R.}~\bibnamefont
  {Ganesh}},\ }\bibfield  {title} {\bibinfo {title} {Disorder-induced currents
  as signatures of chiral superconductivity},\ }\href
  {https://doi.org/10.1103/PhysRevB.104.094505} {\bibfield  {journal} {\bibinfo
   {journal} {Phys. Rev. B}\ }\textbf {\bibinfo {volume} {104}},\ \bibinfo
  {pages} {094505} (\bibinfo {year} {2021})}\BibitemShut {NoStop}%
\bibitem [{\citenamefont {Lee}\ \emph {et~al.}(2009)\citenamefont {Lee},
  \citenamefont {Zhang},\ and\ \citenamefont {Wu}}]{Lee2009}%
  \BibitemOpen
  \bibfield  {author} {\bibinfo {author} {\bibfnamefont {W.-C.}\ \bibnamefont
  {Lee}}, \bibinfo {author} {\bibfnamefont {S.-C.}\ \bibnamefont {Zhang}},\
  and\ \bibinfo {author} {\bibfnamefont {C.}~\bibnamefont {Wu}},\ }\bibfield
  {title} {\bibinfo {title} {Pairing state with a time-reversal symmetry
  breaking in feas-based superconductors},\ }\href
  {https://doi.org/10.1103/PhysRevLett.102.217002} {\bibfield  {journal}
  {\bibinfo  {journal} {Phys. Rev. Lett.}\ }\textbf {\bibinfo {volume} {102}},\
  \bibinfo {pages} {217002} (\bibinfo {year} {2009})}\BibitemShut {NoStop}%
\bibitem [{\citenamefont {Maiti}\ \emph {et~al.}(2015)\citenamefont {Maiti},
  \citenamefont {Sigrist},\ and\ \citenamefont {Chubukov}}]{Maiti2015}%
  \BibitemOpen
  \bibfield  {author} {\bibinfo {author} {\bibfnamefont {S.}~\bibnamefont
  {Maiti}}, \bibinfo {author} {\bibfnamefont {M.}~\bibnamefont {Sigrist}},\
  and\ \bibinfo {author} {\bibfnamefont {A.}~\bibnamefont {Chubukov}},\
  }\bibfield  {title} {\bibinfo {title} {Spontaneous currents in a
  superconductor with $s+is$ symmetry},\ }\href
  {https://doi.org/10.1103/PhysRevB.91.161102} {\bibfield  {journal} {\bibinfo
  {journal} {Phys. Rev. B}\ }\textbf {\bibinfo {volume} {91}},\ \bibinfo
  {pages} {161102} (\bibinfo {year} {2015})}\BibitemShut {NoStop}%
\bibitem [{\citenamefont {Lin}\ \emph {et~al.}(2016)\citenamefont {Lin},
  \citenamefont {Maiti},\ and\ \citenamefont {Chubukov}}]{Lin2016}%
  \BibitemOpen
  \bibfield  {author} {\bibinfo {author} {\bibfnamefont {S.-Z.}\ \bibnamefont
  {Lin}}, \bibinfo {author} {\bibfnamefont {S.}~\bibnamefont {Maiti}},\ and\
  \bibinfo {author} {\bibfnamefont {A.}~\bibnamefont {Chubukov}},\ }\bibfield
  {title} {\bibinfo {title} {Distinguishing between $s+id$ and $s+is$ pairing
  symmetries in multiband superconductors through spontaneous magnetization
  pattern induced by a defect},\ }\href
  {https://doi.org/10.1103/PhysRevB.94.064519} {\bibfield  {journal} {\bibinfo
  {journal} {Phys. Rev. B}\ }\textbf {\bibinfo {volume} {94}},\ \bibinfo
  {pages} {064519} (\bibinfo {year} {2016})}\BibitemShut {NoStop}%
\bibitem [{\citenamefont {Li}\ \emph {et~al.}(2021)\citenamefont {Li},
  \citenamefont {Kivelson},\ and\ \citenamefont {Lee}}]{Li2021}%
  \BibitemOpen
  \bibfield  {author} {\bibinfo {author} {\bibfnamefont {Z.-X.}\ \bibnamefont
  {Li}}, \bibinfo {author} {\bibfnamefont {S.~A.}\ \bibnamefont {Kivelson}},\
  and\ \bibinfo {author} {\bibfnamefont {D.-H.}\ \bibnamefont {Lee}},\
  }\bibfield  {title} {\bibinfo {title} {Superconductor-to-metal transition in
  overdoped cuprates},\ }\href {https://doi.org/10.1038/s41535-021-00335-4}
  {\bibfield  {journal} {\bibinfo  {journal} {npj Quantum Materials}\ }\textbf
  {\bibinfo {volume} {6}},\ \bibinfo {pages} {36} (\bibinfo {year}
  {2021})}\BibitemShut {NoStop}%
\bibitem [{\citenamefont {Hirschfeld}\ \emph {et~al.}(1986)\citenamefont
  {Hirschfeld}, \citenamefont {Vollhardt},\ and\ \citenamefont
  {Wölfle}}]{Hirschfeld1986}%
  \BibitemOpen
  \bibfield  {author} {\bibinfo {author} {\bibfnamefont {P.}~\bibnamefont
  {Hirschfeld}}, \bibinfo {author} {\bibfnamefont {D.}~\bibnamefont
  {Vollhardt}},\ and\ \bibinfo {author} {\bibfnamefont {P.}~\bibnamefont
  {Wölfle}},\ }\bibfield  {title} {\bibinfo {title} {Resonant impurity
  scattering in heavy fermion superconductors},\ }\href
  {https://doi.org/https://doi.org/10.1016/0038-1098(86)90190-0} {\bibfield
  {journal} {\bibinfo  {journal} {Solid State Communications}\ }\textbf
  {\bibinfo {volume} {59}},\ \bibinfo {pages} {111} (\bibinfo {year}
  {1986})}\BibitemShut {NoStop}%
\bibitem [{\citenamefont {Schmitt-Rink}\ \emph {et~al.}(1986)\citenamefont
  {Schmitt-Rink}, \citenamefont {Miyake},\ and\ \citenamefont
  {Varma}}]{SchmittRink1986}%
  \BibitemOpen
  \bibfield  {author} {\bibinfo {author} {\bibfnamefont {S.}~\bibnamefont
  {Schmitt-Rink}}, \bibinfo {author} {\bibfnamefont {K.}~\bibnamefont
  {Miyake}},\ and\ \bibinfo {author} {\bibfnamefont {C.~M.}\ \bibnamefont
  {Varma}},\ }\bibfield  {title} {\bibinfo {title} {Transport and thermal
  properties of heavy-fermion superconductors: A unified picture},\ }\href
  {https://doi.org/10.1103/PhysRevLett.57.2575} {\bibfield  {journal} {\bibinfo
   {journal} {Phys. Rev. Lett.}\ }\textbf {\bibinfo {volume} {57}},\ \bibinfo
  {pages} {2575} (\bibinfo {year} {1986})}\BibitemShut {NoStop}%
\bibitem [{\citenamefont {Hirschfeld}\ \emph {et~al.}(1988)\citenamefont
  {Hirschfeld}, \citenamefont {W\"olfle},\ and\ \citenamefont
  {Einzel}}]{Hirschfeld1988}%
  \BibitemOpen
  \bibfield  {author} {\bibinfo {author} {\bibfnamefont {P.~J.}\ \bibnamefont
  {Hirschfeld}}, \bibinfo {author} {\bibfnamefont {P.}~\bibnamefont
  {W\"olfle}},\ and\ \bibinfo {author} {\bibfnamefont {D.}~\bibnamefont
  {Einzel}},\ }\bibfield  {title} {\bibinfo {title} {Consequences of resonant
  impurity scattering in anisotropic superconductors: Thermal and spin
  relaxation properties},\ }\href {https://doi.org/10.1103/PhysRevB.37.83}
  {\bibfield  {journal} {\bibinfo  {journal} {Phys. Rev. B}\ }\textbf {\bibinfo
  {volume} {37}},\ \bibinfo {pages} {83} (\bibinfo {year} {1988})}\BibitemShut
  {NoStop}%
\bibitem [{\citenamefont {Prohammer}\ and\ \citenamefont
  {Carbotte}(1991)}]{Prohammer1991}%
  \BibitemOpen
  \bibfield  {author} {\bibinfo {author} {\bibfnamefont {M.}~\bibnamefont
  {Prohammer}}\ and\ \bibinfo {author} {\bibfnamefont {J.~P.}\ \bibnamefont
  {Carbotte}},\ }\bibfield  {title} {\bibinfo {title} {London penetration depth
  of d-wave superconductors},\ }\href
  {https://doi.org/10.1103/PhysRevB.43.5370} {\bibfield  {journal} {\bibinfo
  {journal} {Phys. Rev. B}\ }\textbf {\bibinfo {volume} {43}},\ \bibinfo
  {pages} {5370} (\bibinfo {year} {1991})}\BibitemShut {NoStop}%
\bibitem [{\citenamefont {Hirschfeld}\ and\ \citenamefont
  {Goldenfeld}(1993)}]{felds1993}%
  \BibitemOpen
  \bibfield  {author} {\bibinfo {author} {\bibfnamefont {P.~J.}\ \bibnamefont
  {Hirschfeld}}\ and\ \bibinfo {author} {\bibfnamefont {N.}~\bibnamefont
  {Goldenfeld}},\ }\bibfield  {title} {\bibinfo {title} {Effect of strong
  scattering on the low-temperature penetration depth of a d-wave
  superconductor},\ }\href {https://doi.org/10.1103/PhysRevB.48.4219}
  {\bibfield  {journal} {\bibinfo  {journal} {Phys. Rev. B}\ }\textbf {\bibinfo
  {volume} {48}},\ \bibinfo {pages} {4219} (\bibinfo {year}
  {1993})}\BibitemShut {NoStop}%
\bibitem [{\citenamefont {Tsuchiura}\ \emph {et~al.}(2001)\citenamefont
  {Tsuchiura}, \citenamefont {Tanaka}, \citenamefont {Ogata},\ and\
  \citenamefont {Kashiwaya}}]{Tsuchiura2001}%
  \BibitemOpen
  \bibfield  {author} {\bibinfo {author} {\bibfnamefont {H.}~\bibnamefont
  {Tsuchiura}}, \bibinfo {author} {\bibfnamefont {Y.}~\bibnamefont {Tanaka}},
  \bibinfo {author} {\bibfnamefont {M.}~\bibnamefont {Ogata}},\ and\ \bibinfo
  {author} {\bibfnamefont {S.}~\bibnamefont {Kashiwaya}},\ }\bibfield  {title}
  {\bibinfo {title} {Local magnetic moments around a nonmagnetic impurity in
  the two-dimensional $t\ensuremath{-}j$ model},\ }\href
  {https://doi.org/10.1103/PhysRevB.64.140501} {\bibfield  {journal} {\bibinfo
  {journal} {Phys. Rev. B}\ }\textbf {\bibinfo {volume} {64}},\ \bibinfo
  {pages} {140501} (\bibinfo {year} {2001})}\BibitemShut {NoStop}%
\bibitem [{\citenamefont {Wang}\ and\ \citenamefont {Lee}(2002)}]{ZWang2002}%
  \BibitemOpen
  \bibfield  {author} {\bibinfo {author} {\bibfnamefont {Z.}~\bibnamefont
  {Wang}}\ and\ \bibinfo {author} {\bibfnamefont {P.~A.}\ \bibnamefont {Lee}},\
  }\bibfield  {title} {\bibinfo {title} {Local moment formation in the
  superconducting state of a doped mott insulator},\ }\href
  {https://doi.org/10.1103/PhysRevLett.89.217002} {\bibfield  {journal}
  {\bibinfo  {journal} {Phys. Rev. Lett.}\ }\textbf {\bibinfo {volume} {89}},\
  \bibinfo {pages} {217002} (\bibinfo {year} {2002})}\BibitemShut {NoStop}%
\bibitem [{\citenamefont {Zhu}\ \emph {et~al.}(2002)\citenamefont {Zhu},
  \citenamefont {Martin},\ and\ \citenamefont {Bishop}}]{Zhu2002}%
  \BibitemOpen
  \bibfield  {author} {\bibinfo {author} {\bibfnamefont {J.-X.}\ \bibnamefont
  {Zhu}}, \bibinfo {author} {\bibfnamefont {I.}~\bibnamefont {Martin}},\ and\
  \bibinfo {author} {\bibfnamefont {A.~R.}\ \bibnamefont {Bishop}},\ }\bibfield
   {title} {\bibinfo {title} {Spin and charge order around vortices and
  impurities in high-${T}_{c}$ superconductors},\ }\href
  {https://doi.org/10.1103/PhysRevLett.89.067003} {\bibfield  {journal}
  {\bibinfo  {journal} {Phys. Rev. Lett.}\ }\textbf {\bibinfo {volume} {89}},\
  \bibinfo {pages} {067003} (\bibinfo {year} {2002})}\BibitemShut {NoStop}%
\bibitem [{\citenamefont {Chen}\ and\ \citenamefont {Ting}(2004)}]{Chen2004}%
  \BibitemOpen
  \bibfield  {author} {\bibinfo {author} {\bibfnamefont {Y.}~\bibnamefont
  {Chen}}\ and\ \bibinfo {author} {\bibfnamefont {C.~S.}\ \bibnamefont
  {Ting}},\ }\bibfield  {title} {\bibinfo {title} {States of local moment
  induced by nonmagnetic impurities in cuprate superconductors},\ }\href
  {https://doi.org/10.1103/PhysRevLett.92.077203} {\bibfield  {journal}
  {\bibinfo  {journal} {Phys. Rev. Lett.}\ }\textbf {\bibinfo {volume} {92}},\
  \bibinfo {pages} {077203} (\bibinfo {year} {2004})}\BibitemShut {NoStop}%
\bibitem [{\citenamefont {Andersen}\ \emph {et~al.}(2007)\citenamefont
  {Andersen}, \citenamefont {Hirschfeld}, \citenamefont {Kampf},\ and\
  \citenamefont {Schmid}}]{Andersen2007}%
  \BibitemOpen
  \bibfield  {author} {\bibinfo {author} {\bibfnamefont {B.~M.}\ \bibnamefont
  {Andersen}}, \bibinfo {author} {\bibfnamefont {P.~J.}\ \bibnamefont
  {Hirschfeld}}, \bibinfo {author} {\bibfnamefont {A.~P.}\ \bibnamefont
  {Kampf}},\ and\ \bibinfo {author} {\bibfnamefont {M.}~\bibnamefont
  {Schmid}},\ }\bibfield  {title} {\bibinfo {title} {Disorder-induced static
  antiferromagnetism in cuprate superconductors},\ }\href
  {https://doi.org/10.1103/PhysRevLett.99.147002} {\bibfield  {journal}
  {\bibinfo  {journal} {Phys. Rev. Lett.}\ }\textbf {\bibinfo {volume} {99}},\
  \bibinfo {pages} {147002} (\bibinfo {year} {2007})}\BibitemShut {NoStop}%
\bibitem [{\citenamefont {Harter}\ \emph {et~al.}(2007)\citenamefont {Harter},
  \citenamefont {Andersen}, \citenamefont {Bobroff}, \citenamefont {Gabay},\
  and\ \citenamefont {Hirschfeld}}]{Harter2007}%
  \BibitemOpen
  \bibfield  {author} {\bibinfo {author} {\bibfnamefont {J.~W.}\ \bibnamefont
  {Harter}}, \bibinfo {author} {\bibfnamefont {B.~M.}\ \bibnamefont
  {Andersen}}, \bibinfo {author} {\bibfnamefont {J.}~\bibnamefont {Bobroff}},
  \bibinfo {author} {\bibfnamefont {M.}~\bibnamefont {Gabay}},\ and\ \bibinfo
  {author} {\bibfnamefont {P.~J.}\ \bibnamefont {Hirschfeld}},\ }\bibfield
  {title} {\bibinfo {title} {Antiferromagnetic correlations and impurity
  broadening of nmr linewidths in cuprate superconductors},\ }\href
  {https://doi.org/10.1103/PhysRevB.75.054520} {\bibfield  {journal} {\bibinfo
  {journal} {Phys. Rev. B}\ }\textbf {\bibinfo {volume} {75}},\ \bibinfo
  {pages} {054520} (\bibinfo {year} {2007})}\BibitemShut {NoStop}%
\bibitem [{\citenamefont {Andersen}\ \emph {et~al.}(2010)\citenamefont
  {Andersen}, \citenamefont {Graser},\ and\ \citenamefont
  {Hirschfeld}}]{Andersen2010}%
  \BibitemOpen
  \bibfield  {author} {\bibinfo {author} {\bibfnamefont {B.~M.}\ \bibnamefont
  {Andersen}}, \bibinfo {author} {\bibfnamefont {S.}~\bibnamefont {Graser}},\
  and\ \bibinfo {author} {\bibfnamefont {P.~J.}\ \bibnamefont {Hirschfeld}},\
  }\bibfield  {title} {\bibinfo {title} {Disorder-induced freezing of dynamical
  spin fluctuations in underdoped cuprate superconductors},\ }\href
  {https://doi.org/10.1103/PhysRevLett.105.147002} {\bibfield  {journal}
  {\bibinfo  {journal} {Phys. Rev. Lett.}\ }\textbf {\bibinfo {volume} {105}},\
  \bibinfo {pages} {147002} (\bibinfo {year} {2010})}\BibitemShut {NoStop}%
\bibitem [{\citenamefont {Schmid}\ \emph {et~al.}(2010)\citenamefont {Schmid},
  \citenamefont {Andersen}, \citenamefont {Kampf},\ and\ \citenamefont
  {Hirschfeld}}]{Schmid_2010}%
  \BibitemOpen
  \bibfield  {author} {\bibinfo {author} {\bibfnamefont {M.}~\bibnamefont
  {Schmid}}, \bibinfo {author} {\bibfnamefont {B.~M.}\ \bibnamefont
  {Andersen}}, \bibinfo {author} {\bibfnamefont {A.~P.}\ \bibnamefont
  {Kampf}},\ and\ \bibinfo {author} {\bibfnamefont {P.~J.}\ \bibnamefont
  {Hirschfeld}},\ }\bibfield  {title} {\bibinfo {title} {d-wave
  superconductivity as a catalyst for antiferromagnetism in underdoped
  cuprates},\ }\href {https://doi.org/10.1088/1367-2630/12/5/053043} {\bibfield
   {journal} {\bibinfo  {journal} {New Journal of Physics}\ }\textbf {\bibinfo
  {volume} {12}},\ \bibinfo {pages} {053043} (\bibinfo {year}
  {2010})}\BibitemShut {NoStop}%
\bibitem [{\citenamefont {Gastiasoro}\ \emph {et~al.}(2013)\citenamefont
  {Gastiasoro}, \citenamefont {Hirschfeld},\ and\ \citenamefont
  {Andersen}}]{Gastiasoro2013}%
  \BibitemOpen
  \bibfield  {author} {\bibinfo {author} {\bibfnamefont {M.~N.}\ \bibnamefont
  {Gastiasoro}}, \bibinfo {author} {\bibfnamefont {P.~J.}\ \bibnamefont
  {Hirschfeld}},\ and\ \bibinfo {author} {\bibfnamefont {B.~M.}\ \bibnamefont
  {Andersen}},\ }\bibfield  {title} {\bibinfo {title} {Impurity states and
  cooperative magnetic order in fe-based superconductors},\ }\href
  {https://doi.org/10.1103/PhysRevB.88.220509} {\bibfield  {journal} {\bibinfo
  {journal} {Phys. Rev. B}\ }\textbf {\bibinfo {volume} {88}},\ \bibinfo
  {pages} {220509} (\bibinfo {year} {2013})}\BibitemShut {NoStop}%
\bibitem [{\citenamefont {Gastiasoro}\ \emph {et~al.}(2014)\citenamefont
  {Gastiasoro}, \citenamefont {Paul}, \citenamefont {Wang}, \citenamefont
  {Hirschfeld},\ and\ \citenamefont {Andersen}}]{Gastiasoro2014}%
  \BibitemOpen
  \bibfield  {author} {\bibinfo {author} {\bibfnamefont {M.~N.}\ \bibnamefont
  {Gastiasoro}}, \bibinfo {author} {\bibfnamefont {I.}~\bibnamefont {Paul}},
  \bibinfo {author} {\bibfnamefont {Y.}~\bibnamefont {Wang}}, \bibinfo {author}
  {\bibfnamefont {P.~J.}\ \bibnamefont {Hirschfeld}},\ and\ \bibinfo {author}
  {\bibfnamefont {B.~M.}\ \bibnamefont {Andersen}},\ }\bibfield  {title}
  {\bibinfo {title} {Emergent defect states as a source of resistivity
  anisotropy in the nematic phase of iron pnictides},\ }\href
  {https://doi.org/10.1103/PhysRevLett.113.127001} {\bibfield  {journal}
  {\bibinfo  {journal} {Phys. Rev. Lett.}\ }\textbf {\bibinfo {volume} {113}},\
  \bibinfo {pages} {127001} (\bibinfo {year} {2014})}\BibitemShut {NoStop}%
\bibitem [{\citenamefont {Gastiasoro}\ and\ \citenamefont
  {Andersen}(2015)}]{Gastiasoro2015}%
  \BibitemOpen
  \bibfield  {author} {\bibinfo {author} {\bibfnamefont {M.~N.}\ \bibnamefont
  {Gastiasoro}}\ and\ \bibinfo {author} {\bibfnamefont {B.~M.}\ \bibnamefont
  {Andersen}},\ }\bibfield  {title} {\bibinfo {title} {Local magnetization
  nucleated by non-magnetic impurities in fe-based superconductors},\ }\href
  {https://doi.org/10.1007/s10948-014-2908-2} {\bibfield  {journal} {\bibinfo
  {journal} {Journal of Superconductivity and Novel Magnetism}\ }\textbf
  {\bibinfo {volume} {28}},\ \bibinfo {pages} {1321} (\bibinfo {year}
  {2015})}\BibitemShut {NoStop}%
\bibitem [{\citenamefont {Martiny}\ \emph {et~al.}(2015)\citenamefont
  {Martiny}, \citenamefont {Gastiasoro}, \citenamefont {Vekhter},\ and\
  \citenamefont {Andersen}}]{Martiny2015}%
  \BibitemOpen
  \bibfield  {author} {\bibinfo {author} {\bibfnamefont {J.~H.~J.}\
  \bibnamefont {Martiny}}, \bibinfo {author} {\bibfnamefont {M.~N.}\
  \bibnamefont {Gastiasoro}}, \bibinfo {author} {\bibfnamefont
  {I.}~\bibnamefont {Vekhter}},\ and\ \bibinfo {author} {\bibfnamefont {B.~M.}\
  \bibnamefont {Andersen}},\ }\bibfield  {title} {\bibinfo {title}
  {Impurity-induced antiferromagnetic order in pauli-limited nodal
  superconductors: Application to heavy-fermion ${\mathrm{cecoin}}_{5}$},\
  }\href {https://doi.org/10.1103/PhysRevB.92.224510} {\bibfield  {journal}
  {\bibinfo  {journal} {Phys. Rev. B}\ }\textbf {\bibinfo {volume} {92}},\
  \bibinfo {pages} {224510} (\bibinfo {year} {2015})}\BibitemShut {NoStop}%
\bibitem [{\citenamefont {Gastiasoro}\ \emph {et~al.}(2016)\citenamefont
  {Gastiasoro}, \citenamefont {Bernardini},\ and\ \citenamefont
  {Andersen}}]{Gastiasoro2016}%
  \BibitemOpen
  \bibfield  {author} {\bibinfo {author} {\bibfnamefont {M.~N.}\ \bibnamefont
  {Gastiasoro}}, \bibinfo {author} {\bibfnamefont {F.}~\bibnamefont
  {Bernardini}},\ and\ \bibinfo {author} {\bibfnamefont {B.~M.}\ \bibnamefont
  {Andersen}},\ }\bibfield  {title} {\bibinfo {title} {Unconventional disorder
  effects in correlated superconductors},\ }\href
  {https://doi.org/10.1103/PhysRevLett.117.257002} {\bibfield  {journal}
  {\bibinfo  {journal} {Phys. Rev. Lett.}\ }\textbf {\bibinfo {volume} {117}},\
  \bibinfo {pages} {257002} (\bibinfo {year} {2016})}\BibitemShut {NoStop}%
\bibitem [{\citenamefont {Martiny}\ \emph {et~al.}(2019)\citenamefont
  {Martiny}, \citenamefont {Kreisel},\ and\ \citenamefont
  {Andersen}}]{Martiny2019}%
  \BibitemOpen
  \bibfield  {author} {\bibinfo {author} {\bibfnamefont {J.~H.~J.}\
  \bibnamefont {Martiny}}, \bibinfo {author} {\bibfnamefont {A.}~\bibnamefont
  {Kreisel}},\ and\ \bibinfo {author} {\bibfnamefont {B.~M.}\ \bibnamefont
  {Andersen}},\ }\bibfield  {title} {\bibinfo {title} {Theoretical study of
  impurity-induced magnetism in fese},\ }\href
  {https://doi.org/10.1103/PhysRevB.99.014509} {\bibfield  {journal} {\bibinfo
  {journal} {Phys. Rev. B}\ }\textbf {\bibinfo {volume} {99}},\ \bibinfo
  {pages} {014509} (\bibinfo {year} {2019})}\BibitemShut {NoStop}%
\bibitem [{\citenamefont {R\o{}mer}\ \emph {et~al.}(2015)\citenamefont
  {R\o{}mer}, \citenamefont {Kreisel}, \citenamefont {Eremin}, \citenamefont
  {Malakhov}, \citenamefont {Maier}, \citenamefont {Hirschfeld},\ and\
  \citenamefont {Andersen}}]{Romer2015}%
  \BibitemOpen
  \bibfield  {author} {\bibinfo {author} {\bibfnamefont {A.~T.}\ \bibnamefont
  {R\o{}mer}}, \bibinfo {author} {\bibfnamefont {A.}~\bibnamefont {Kreisel}},
  \bibinfo {author} {\bibfnamefont {I.}~\bibnamefont {Eremin}}, \bibinfo
  {author} {\bibfnamefont {M.~A.}\ \bibnamefont {Malakhov}}, \bibinfo {author}
  {\bibfnamefont {T.~A.}\ \bibnamefont {Maier}}, \bibinfo {author}
  {\bibfnamefont {P.~J.}\ \bibnamefont {Hirschfeld}},\ and\ \bibinfo {author}
  {\bibfnamefont {B.~M.}\ \bibnamefont {Andersen}},\ }\bibfield  {title}
  {\bibinfo {title} {Pairing symmetry of the one-band hubbard model in the
  paramagnetic weak-coupling limit: A numerical rpa study},\ }\href
  {https://doi.org/10.1103/PhysRevB.92.104505} {\bibfield  {journal} {\bibinfo
  {journal} {Phys. Rev. B}\ }\textbf {\bibinfo {volume} {92}},\ \bibinfo
  {pages} {104505} (\bibinfo {year} {2015})}\BibitemShut {NoStop}%
\bibitem [{\citenamefont {Kreisel}\ \emph {et~al.}(2017)\citenamefont
  {Kreisel}, \citenamefont {R{\o}mer}, \citenamefont {Hirschfeld},\ and\
  \citenamefont {Andersen}}]{Kreisel2017}%
  \BibitemOpen
  \bibfield  {author} {\bibinfo {author} {\bibfnamefont {A.}~\bibnamefont
  {Kreisel}}, \bibinfo {author} {\bibfnamefont {A.~T.}\ \bibnamefont
  {R{\o}mer}}, \bibinfo {author} {\bibfnamefont {P.~J.}\ \bibnamefont
  {Hirschfeld}},\ and\ \bibinfo {author} {\bibfnamefont {B.~M.}\ \bibnamefont
  {Andersen}},\ }\bibfield  {title} {\bibinfo {title} {Superconducting phase
  diagram of the paramagnetic one-band hubbard model},\ }\href
  {https://doi.org/10.1007/s10948-016-3758-x} {\bibfield  {journal} {\bibinfo
  {journal} {Journal of Superconductivity and Novel Magnetism}\ }\textbf
  {\bibinfo {volume} {30}},\ \bibinfo {pages} {85} (\bibinfo {year}
  {2017})}\BibitemShut {NoStop}%
\bibitem [{\citenamefont {Garaud}\ and\ \citenamefont
  {Babaev}(2014)}]{Garaud2014}%
  \BibitemOpen
  \bibfield  {author} {\bibinfo {author} {\bibfnamefont {J.}~\bibnamefont
  {Garaud}}\ and\ \bibinfo {author} {\bibfnamefont {E.}~\bibnamefont
  {Babaev}},\ }\bibfield  {title} {\bibinfo {title} {Domain walls and their
  experimental signatures in $s+is$ superconductors},\ }\href
  {https://doi.org/10.1103/PhysRevLett.112.017003} {\bibfield  {journal}
  {\bibinfo  {journal} {Phys. Rev. Lett.}\ }\textbf {\bibinfo {volume} {112}},\
  \bibinfo {pages} {017003} (\bibinfo {year} {2014})}\BibitemShut {NoStop}%
\bibitem [{\citenamefont {Andersen}\ \emph {et~al.}(2006)\citenamefont
  {Andersen}, \citenamefont {Melikyan}, \citenamefont {Nunner},\ and\
  \citenamefont {Hirschfeld}}]{Phase_impurities}%
  \BibitemOpen
  \bibfield  {author} {\bibinfo {author} {\bibfnamefont {B.~M.}\ \bibnamefont
  {Andersen}}, \bibinfo {author} {\bibfnamefont {A.}~\bibnamefont {Melikyan}},
  \bibinfo {author} {\bibfnamefont {T.~S.}\ \bibnamefont {Nunner}},\ and\
  \bibinfo {author} {\bibfnamefont {P.~J.}\ \bibnamefont {Hirschfeld}},\
  }\bibfield  {title} {\bibinfo {title} {Andreev states near short-ranged
  pairing potential impurities},\ }\href
  {https://doi.org/10.1103/PhysRevLett.96.097004} {\bibfield  {journal}
  {\bibinfo  {journal} {Phys. Rev. Lett.}\ }\textbf {\bibinfo {volume} {96}},\
  \bibinfo {pages} {097004} (\bibinfo {year} {2006})}\BibitemShut {NoStop}%
\bibitem [{\citenamefont {Grinenko}\ \emph {et~al.}(2020)\citenamefont
  {Grinenko}, \citenamefont {Sarkar}, \citenamefont {Kihou}, \citenamefont
  {Lee}, \citenamefont {Morozov}, \citenamefont {Aswartham}, \citenamefont
  {B{\"u}chner}, \citenamefont {Chekhonin}, \citenamefont {Skrotzki},
  \citenamefont {Nenkov}, \citenamefont {H{\"u}hne}, \citenamefont {Nielsch},
  \citenamefont {Drechsler}, \citenamefont {Vadimov}, \citenamefont {Silaev},
  \citenamefont {Volkov}, \citenamefont {Eremin}, \citenamefont {Luetkens},\
  and\ \citenamefont {Klauss}}]{Grinenko2020}%
  \BibitemOpen
  \bibfield  {author} {\bibinfo {author} {\bibfnamefont {V.}~\bibnamefont
  {Grinenko}}, \bibinfo {author} {\bibfnamefont {R.}~\bibnamefont {Sarkar}},
  \bibinfo {author} {\bibfnamefont {K.}~\bibnamefont {Kihou}}, \bibinfo
  {author} {\bibfnamefont {C.~H.}\ \bibnamefont {Lee}}, \bibinfo {author}
  {\bibfnamefont {I.}~\bibnamefont {Morozov}}, \bibinfo {author} {\bibfnamefont
  {S.}~\bibnamefont {Aswartham}}, \bibinfo {author} {\bibfnamefont
  {B.}~\bibnamefont {B{\"u}chner}}, \bibinfo {author} {\bibfnamefont
  {P.}~\bibnamefont {Chekhonin}}, \bibinfo {author} {\bibfnamefont
  {W.}~\bibnamefont {Skrotzki}}, \bibinfo {author} {\bibfnamefont
  {K.}~\bibnamefont {Nenkov}}, \bibinfo {author} {\bibfnamefont
  {R.}~\bibnamefont {H{\"u}hne}}, \bibinfo {author} {\bibfnamefont
  {K.}~\bibnamefont {Nielsch}}, \bibinfo {author} {\bibfnamefont {S.-L.}\
  \bibnamefont {Drechsler}}, \bibinfo {author} {\bibfnamefont {V.~L.}\
  \bibnamefont {Vadimov}}, \bibinfo {author} {\bibfnamefont {M.~A.}\
  \bibnamefont {Silaev}}, \bibinfo {author} {\bibfnamefont {P.~A.}\
  \bibnamefont {Volkov}}, \bibinfo {author} {\bibfnamefont {I.}~\bibnamefont
  {Eremin}}, \bibinfo {author} {\bibfnamefont {H.}~\bibnamefont {Luetkens}},\
  and\ \bibinfo {author} {\bibfnamefont {H.-H.}\ \bibnamefont {Klauss}},\
  }\bibfield  {title} {\bibinfo {title} {Superconductivity with broken
  time-reversal symmetry inside a superconducting s-wave state},\ }\href
  {https://doi.org/10.1038/s41567-020-0886-9} {\bibfield  {journal} {\bibinfo
  {journal} {Nature Physics}\ }\textbf {\bibinfo {volume} {16}},\ \bibinfo
  {pages} {789} (\bibinfo {year} {2020})}\BibitemShut {NoStop}%
\end{thebibliography}%
\end{document}